\documentclass[12pt]{article}
\usepackage[utf8]{inputenc}
\usepackage[]{graphicx}
\usepackage{epstopdf}
\usepackage{epsfig}
\usepackage{amsmath}
\usepackage{amssymb}
\usepackage{subcaption}
\usepackage{color}
\usepackage{verbatim}
\usepackage{multirow}
\usepackage[bottom]{footmisc}
\usepackage{cite,url}
\usepackage{hyperref}
\usepackage[normalem]{ulem}
\usepackage{xspace}

\evensidemargin - .5truecm
\usepackage{xcolor}

\newcommand{\noun}[1]{{\scshape #1}}
\newcommand{\POWHEG}{\noun{Powheg}\xspace}
\newcommand{\POWHEGBOX}{\noun{Powheg-Box}\xspace}
\newcommand{\ggHH}{\noun{ggHH}\xspace}

\newcommand{\be}{\begin{equation}}
\newcommand{\ee}{\end{equation}}
\newcommand{\nn}{\nonumber}
\newcommand{\bea}{\begin{eqnarray}}
\newcommand{\eea}{\end{eqnarray}}
\newcommand{\bfig}{\begin{figure}}
\newcommand{\efig}{\end{figure}}
\newcommand{\bc}{\begin{center}}
\newcommand{\ec}{\end{center}}

\newcommand{\ktre}{\kappa_{\lambda}}

\def\as{\alpha_s}

\def\sq2{\sqrt{2}}

\newcommand{\smallh}{{\scriptscriptstyle H}}
\newcommand{\mh}{m_\smallh}
\newcommand{\mt}{m_t}

\newcommand{\pt}{p_{\scriptscriptstyle T}}

\newcommand{\tril}{\lambda_{3}}
\newcommand{\trilsm}{\tril^{\rm SM}}
\newcommand{\qual}{\lambda_{4}}
\newcommand\Phid{\Phi^\dagger}

\newcommand{\MSbar}{\overline{\mathrm{MS}}}

\begin{document}
\begin{titlepage}
\nopagebreak
{\flushright{
    \begin{minipage}{5cm}
      CERN-TH-2023-131
    \end{minipage}
  }
}
\renewcommand{\thefootnote}{\fnsymbol{footnote}}
\vspace{1cm}
\begin{center}
  {\Large \bf \color{magenta} Higgs boson pair production at NLO in the \POWHEG approach and the top quark mass uncertainties}

\bigskip\color{black}\vspace{0.6cm}
{\large\bf  Emanuele Bagnaschi$^{a,b}$\footnote{email: emanuele.bagnaschi@cern.ch},
      Giuseppe Degrassi$^{c}$\footnote{email: giuseppe.degrassi@uniroma3.it},
       Ramona Gr\"{o}ber$^{d}$\footnote{email: ramona.groeber@pd.infn.it}}
     \\[7mm]
{\it  (a) Istituto Nazionale di Fisica Nucleare, Laboratori Nazionali di Frascati, C.P. 13, 00044 Frascati, Italy}\\[1mm]
{\it  (b) CERN, Theoretical Physics Department, 1211 Geneva 23,
         Switzerland}\\[1mm]
{\it (c) Dipartimento di Matematica e Fisica, Universit{\`a} di Roma Tre and \\
  INFN, sezione di Roma Tre, I-00146 Rome, Italy}\\[1mm]
{\it (d) Dipartimento di Fisica e Astronomia 'G.~Galilei', Universit\`a di Padova and INFN, sezione di Padova, I-35131 Padova, Italy}\\
\end{center}

\bigskip
\bigskip
\bigskip
\vspace{0.cm}

\begin{abstract}
  We present a new Monte Carlo code for Higgs boson pair production at
  next-to-leading order in the \POWHEGBOX Monte Carlo approach. The code is
  based on analytic results for the two loop virtual corrections
  which include the full top quark mass dependence. This feature allows to
  freely assign the value of all input parameters, including the trilinear Higgs
  boson self coupling, as well as to vary the
  renormalization scheme employed  for the top quark mass.
  We study the uncertainties due to the top-mass renormalization scheme
  allowing the trilinear Higgs boson self coupling to vary around its Standard
  Model value including parton shower effects.
  Results are  presented for both inclusive and differential observables.
\end{abstract}
\vfill
\end{titlepage}

\setcounter{footnote}{0}

\section{Introduction}

With the discovery of the Higgs boson \cite{ATLAS:2012yve, CMS:2012qbp}
the study of its potential and 
self-interactions has become of great interest to the scientific community
as an ultimate probe of the mechanism of electroweak symmetry breaking.

In the Standard Model (SM), the Higgs potential in the unitary gauge reads
\be
V(H) =  \frac{\mh^2}{2} H^2 + \tril v  H^3 +
\frac{\qual}4 H^4 ,  \label{eq:potun}
\ee
where the Higgs mass ($\mh$) and the trilinear $(\tril)$ and quartic $(\qual)$
interactions are linked by the relations
$\qual^{\rm SM}=\trilsm = \lambda =\mh^2/(2\,v^2)$,
where $v=(\sqrt2 \,G_\mu)^{-1/2}$ is the vacuum expectation value, $G_\mu$ the
Fermi constant and  $\lambda$ is the coefficient of the $(\Phid \Phi)^2$
interaction, $\Phi$ being the Higgs doublet field.

The Higgs mass has now been measured at the
level of $~$one per mille precision \cite{ATLAS:2022net,CMS:2020xrn}.
The trilinear Higgs self-coupling is hence predicted within the SM and its
  determination at the Large Hadron Collider (LHC), where it
  is accessible  via the production of Higgs boson pairs,  thus provides a probe
  of the SM.
However, the main production mode, gluon fusion, has a very small SM
cross-section \cite{DiMicco:2019ngk} and sensitivity to the
$\tril$ SM value has not yet been reached, so that only constraints on $\tril$
can be derived for now. However, during the recent years, the interval of
allowed values for
$\tril$ has shrunk significantly and this trend will continue  during  Run 3
and the High-Luminosity (HL) phase of the LHC. Presently, from the analyses of
the decay channels,
$HH \to b \bar{b} \gamma \gamma, \: HH \to b \bar{b} \tau \tau$ and
$HH \to b \bar{b} b \bar{b}$, the ATLAS Collaboration excluded values outside
the interval
$ -0.6 < \ktre < 6.6$, where $\ktre = \tril/\trilsm$,  at 95\%
confidence level (CL) under the assumption that all the other couplings have SM
values
\cite{ATLAS:2022kbf}.\footnote{The CMS Collaboration reported a bound
$-1.2 <\ktre < 6.5$
slightly weaker \cite{CMS:2022dwd}.} A slightly stronger limit can be
obtained by combining
the information from double Higgs production with the one coming from other
processes that are sensitive to $\tril$ via next-to-leading order (NLO)
electroweak (EW) corrections
\cite{Degrassi:2016wml,Gorbahn:2016uoy,Bizon:2016wgr,
  Maltoni:2017ims,Maltoni:2018ttu,Gorbahn:2019lwq,
  Degrassi:2019yix,Degrassi:2017ucl, Kribs:2017znd, Degrassi:2021uik,
  DiVita:2017eyz, Alasfar:2022zyr}.
Indeed the combination of the information coming from double-Higgs and
single-Higgs production yields $ -0.4 < \ktre < 6.3$ at 95\% CL
\cite{ATLAS:2022kbf}.

In view of future improvements in the experimental analyses of the Higgs pair
production process it is interesting to reappraise the uncertainties in the
theoretical prediction of this process. The process is mediated by heavy quarks
loops that appear in different topologies: i)  the ``signal'' topology
given by triangular diagrams which represent the gluon-fusion production of
an off-shell Higgs that then subsequently decays via the trilinear coupling into
two on-shell bosons and are therefore sensitive to $\tril$; ii)
the ``background'' topology that  at the leading order
(LO) is given by box-like diagrams that do not depend on $\tril$. The LO
calculation of the Higgs boson pair
production has been performed more than thirty years ago
\cite{Glover:1987nx,Dicus:1987ic,Plehn:1996wb}.  The NLO QCD corrections
were first computed in the infinite top mass, $\mt$, limit (HTL)
\cite{Dawson:1998py}, reweighted by the exact
Born amplitude.
Later they were supplemented by the inclusion of $(1/\mt^2)^n$ corrections up
to $n=6$
\cite{Grigo:2013rya,Grigo:2015dia,Degrassi:2016vss}. Then the HTL evaluation
was retained only in the virtual two-loop corrections, while the real ones were
computed including the full top mass
dependence \cite{Frederix:2014hta,Maltoni:2014eza}. Finally, the full top
mass dependence in the virtual correction was obtained by numerical
methods \cite{Borowka:2016ehy,Borowka:2016ypz,Baglio:2018lrj,Baglio:2020ini}.
At the same time  analytic results for the two-loop virtual corrections, valid
in specific regions of the phase space, were presented \cite{Bonciani:2018omm,
  Davies:2018ood,Davies:2018qvx,Wang:2020nnr}.
The analytic evaluation of the corrections via a high-energy (HE) expansion
was later used to replace the numerical evaluation of
Refs.~\cite{Borowka:2016ehy,Borowka:2016ypz} in the high-energy region
in order to  better cover that energy range \cite{Davies:2019dfy}.
Later the HE evaluation was also merged with the analytic evaluation of
the corrections via a Higgs transverse momomentum, $\pt$,  
in order to obtain an analytic result that covers the entire phase
space \cite{Bellafronte:2022jmo}.

The NLO fixed order calculation with the full top mass dependence was also
matched to parton shower programs \cite{Heinrich:2017kxx,Jones:2017giv} and
combined with next-to-next-to-leading order (NNLO) QCD corrections in the
HTL limit \cite{deFlorian:2013uza, deFlorian:2013jea, Grigo:2014jma,
  deFlorian:2016uhr}  while keeping the double real corrections in full top
mass dependence \cite{Grazzini:2018bsd}.
Results at $\text{N}^3\text{LO}$ are available in the HTL \cite{Chen:2019lzz,
Chen:2019fhs}.

The aim of this paper is twofold. On the one side, we present a new
Monte Carlo (MC) code for Higgs pair production in the \POWHEGBOX approach
\cite{Frixione:2007vw,Alioli:2010xd}.
The code retains the full top quark mass dependence at NLO and
is flexible  in the input parameters. In our code
it is possible to vary at the same time both the top mass scheme used
and the value of $\tril$. This is achieved employing an analytic
result for the virtual two-loop corrections instead of a numerical
grid as in the MC code of Refs.~\cite{Heinrich:2017kxx,Heinrich:2019bkc}. On the
other side, we
study the uncertainty related to
the renormalization scheme used for the top mass for arbitrary values
of the trilinear coupling including parton shower effects.

With respect to previous works  in the literature where similar analyses were
presented our work contains several improvements.
In Ref.~\cite{Heinrich:2019bkc}  the dependence on $\tril$ of the
Higgs pair total cross section and differential distributions  was studied
including parton shower effects.  We improve this analysis by addressing also
the dependence  on the top mass renormalization scheme.
 The uncertainties
in double Higgs production related to the top mass scheme were
analyzed in Refs.~\cite{Baglio:2018lrj,Baglio:2020ini,Baglio:2020wgt}.
With respect to these studies, that are based on a NLO fixed order
calculation, we improved the analysis by including parton shower
effects. Furthermore, these previous studies were concentrating on the
investigation of the total cross section or the differential cross
section as a function of the Higgs-pair invariant mass, $M_{HH}$, while we have
the possibility to study other differential distributions.

In this paper we do not discuss the technical details of our MC code. They
can be found in the instruction manual  in the Docs
directory of the source code tree. The code will be
available on the
\POWHEGBOX repository at {\tt https://powhegbox.mib.infn.it}.

The paper is organized as follows. In section \ref{sec2} we describe the
basic feature of our \POWHEG implementation of the $gg \to HH$ process.
In section \ref{sec:res} we present our results for the inclusive cross section
and differential distributions using different top-quark-mass renormalization
schemes and for several values of $\tril$. We will also
compare our results with those of previous analyses. Finally we
present our conclusions.

\section{\POWHEG implementation of $gg \rightarrow HH$}
\label{sec2}
In this section we briefly discuss the main characteristics of our
implementation
of the gluon-fusion Higgs pair production process in the \POWHEGBOX framework.
We first briefly recall the basic features of the \POWHEG formalism, and
the required elements to implement a process in the \POWHEGBOX framework.
We then discuss in more detail our implementation of the virtual
two-loop corrections and of the real radiation contributions.
The way  how these two contributions are implemented is the main difference
between our MC code and the
one presented in Refs. \cite{Heinrich:2017kxx,Heinrich:2019bkc}.

\subsection{The \POWHEG approach}
\label{subsecpowheg}
The \POWHEG formula to match NLO-QCD accurate calculations with parton showers
can be written in a sufficiently
general way as
\begin{align}
  d\sigma = \bar{B}^s(\Phi_B) d\Phi_B \left\{ \Delta^s_{t_0} + \Delta^s_t \frac{R^s(\Phi)}{B(\Phi_B)} d\Phi_r  \right\} + R^f (\Phi) d\Phi + R^{\text{reg}}(\Phi) d\Phi.
  \label{eq:sec2:powheg}
\end{align}
The Born squared matrix element is represented as $B(\Phi_B)$, with
$\Phi_B$ being the phase space at the leading order.  The squared
matrix elements of the real emission, i.e. channels with an additional
parton with respect to the Born process, can be divided in two sets,
according to whether they feature soft/collinear divergences, in which
case we denote them as $R^{\text{div}}(\Phi)$, or not,
$R^{\text{reg}}(\Phi)$. $\Phi$ represents the product of the Born and
the real emission phase spaces $\Phi = \Phi_B \Phi_r$.
In the \POWHEGBOX framework it is possible to split the contribution from the
divergent processes $R^{\text{div}}(\Phi)$ into two terms, $R^s(\Phi)$
and $R^f(\Phi)$.  The term $R^s$ should contain the singular terms,
and it is matched with the parton shower using the \POWHEG Sudakov
form factor $\Delta^s_t$. The definition of the latter is

\begin{align}
  \Delta^s_{t} = e^{-\int \frac{dt'}{t'} \frac{R^s}{B} d\Phi_r \theta(t'-t)}\, ,
\label{eq:sudakov}
\end{align}

with $t$ being the shower ordering variable. The $t_0$ appearing in
eq.~(\ref{eq:sec2:powheg}) is a lower-scale cutoff.  On the other hand,
$R^f$, which should be finite, is simply added as shown in
eq.~(\ref{eq:sec2:powheg}), without any further treatment.  The
arbitrariness in choosing $R^f$ can be used to study the theoretical
uncertainties linked to the matching procedure, since different
definitions differ by higher-order terms.  The non-divergent channels,
$R^{\text{reg}}$, are also added without being multiplied by the
\POWHEG Sudakov form factor.
Finally, the $\bar{B}^s(\Phi_B)$ is the NLO normalization factor
\begin{align}
  \bar{B}^s(\Phi_B) = B(\Phi_B) + \hat{V}_{\text{fin}} (\Phi_B) + \int \hat{R}^s (\Phi_B,\Phi_r) d\Phi_r\, .
\end{align}
In this formula, $\hat{V}_{\text{fin}}$ represents IR- and
UV-regularized two-loop virtual contribution, while $\hat{R}^s$ is the
IR-subtracted real contribution as defined above.

In our MC code $B(\Phi_B)$ is obtained from Ref. \cite{Glover:1987nx}
where the LO amplitude is presented in terms of the Passarino-Veltman functions
\cite{Passarino:1978jh}.  The evaluation of the latters is performed using
the {\tt COLLIER} code \cite{Denner:2014gla}.

\subsection{Virtual two-loop contribution}
\label{subsec2virt}
The virtual two-loop diagrams, that enter in $\hat{V}_{\text{fin}} (\Phi_B)$,
can be assigned to ``signal'' and ``background''
topologies as in the LO case. The contribution of the ``signal'' diagrams
(triangular topology) is known analytically including the full top mass
dependence adapting results for the production of a single Higgs with virtuality
$M_{HH}$ \cite{Aglietti:2006tp,Anastasiou:2006hc}. In our code we implement
the expressions of Ref.~\cite{Aglietti:2006tp}. The possibility of varying the
trilinear coupling is introduced via an additional parameter that rescales the
``signal'' contribution.

The ``background'' topologies are
of two types: double-triangle and box diagrams. Concerning the former,
the diagram topology  is the product of two one-loop triangle diagrams and we
implement in our code the analytic results derived in
Ref.\cite{Degrassi:2016vss} that retain the full top mass dependence.
The box diagrams are the most difficult contribution to evaluate.
The box diagrams depend
on four  energy scales, namely $\hat{s},\, \hat{t},\, \mt, \mh$,
where $\hat{s},\, \hat{t}$, and $\hat{u}$ are the
Mandelstam variables which satisfy the condition
\be
\hat{s} + \hat{t} + \hat{u} = 2 \,\mh^2 ~.
\ee
Alternatively, the scale $\hat{t}$ can be replaced by the transverse momentum
of the Higgs particle, $\pt$.
Exact analytic results for two-loop box diagrams with several energy scales
is at the verge of what can be obtained   with the present computational
technology.  However, using the method of the  expansion of the diagrams in
terms of ratios of small
energy scales vs.~large energy scales, it is possible to obtain an analytic
evaluation of these diagrams
that is valid in specific regions of the phase space where an
hierarchy among the various energy scales present in the diagrams is realized.
The method of the expansion in terms
of the transverse momentum of the Higgs boson \cite{Bonciani:2018omm}
is valid in phase-space regions where $ | \hat{t} |/(4 \mt^2) \lesssim 1$ while
the high-energy (HE) expansion method  \cite{Davies:2018qvx} covers
the complementary regions of the phase space where
$ | \hat{t} |/(4 \mt^2) \gtrsim 1$. In the $\pt$ expansion  it is assumed that
the scales associated
to $m_H$ and to $\pt$ are small compared to the scales set by
$\hat{s}$ and $\mt$. Under this assumption, the box integrals are
expanded in ratios of small over large scales, and the resulting
simplified integrals are written as linear combinations of 52 master
integrals (MI) using Integration-by-Parts (IBP) identities obtained
with $\texttt{LiteRed}$ \cite{Lee:2013mka, Lee:2012cn}. Among the 52 MI,
fifty can be expressed in terms of generalised harmonic polylogarithms while
two are elliptic integrals \cite{vonManteuffel:2017hms}.
On the other hand, in
the HE expansion the two-loop
box integrals are first expanded in terms of small  $\mh$,
then an IBP reduction is performed on the expanded integrals; the
resulting MIs are further expanded in the limit $\mt^2 \ll \hat{s},
|\hat{t}|$ and expressed in terms of harmonic polylogarithms.

As shown in Ref.~\cite{Bellafronte:2022jmo},
in order to merge the two analytic
approximations the  fixed-order results both in the
$\pt$ expansion and in the HE expansion have to be extended up to or beyond
their border of validity, i.e.~$\hat{t} \simeq 4 \mt^2$,  by constructing
a [1/1] Pad\'e  approximant for the $\pt$ expanded result and a [6/6] Pad\'e
approximant for the HE-result. This procedure
reproduces the numerical values \cite{GitHub} in the grid of
Ref.~\cite{Davies:2019dfy},
which is implemented in the MC code presented in
Refs.~\cite{Heinrich:2017kxx,Heinrich:2019bkc}, with an accuracy below the
$1\%$ level.

In our MC code the two-loop box contribution is implemented via the analytic
expressions of  the first three terms
in the $\pt$--expansion and of the first thirteen  terms in the HE--expansion.
With the former terms we construct the [1,1] $\pt$--Pad\'e approximant whose
expression is evaluated when a point in the phase
space satisfies $ | \hat{t} |/(4 \mt^2) < 1$ or $ | \hat{u} |/(4 \mt^2) < 1$.
With the latter
terms we construct a [6,6] HE--Pad\'e approximant whose expression is evaluated
when a point in the phase space  lies in the complementary region,
$ | \hat{t} |/(4 \mt^2) \geq 1$ and $ | \hat{u} |/(4 \mt^2) > 1$.
The evaluation of the (generalised) harmonic polylogarithms is done using
the code \texttt{handyG} \cite{Naterop:2019xaf}, while
the elliptic integrals are evaluated using the routines of
Ref.~\cite{Bonciani:2018uvv}.

Finally, our analytic expressions allow to easily change the
renormalization scheme employed for the top mass as discussed in
Ref.\cite{Bellafronte:2022jmo}.  Our results are presented in the
on-shell (OS) and the modified minimal subtraction ($\MSbar$) top-mass scheme.
In particular, for evaluating the top mass in the $\MSbar$ scheme, we
first convert the OS mass to $m_t^{\MSbar}(\mu_t=m_t^{\textrm{OS}})$
using the two-loop relation\cite{Melnikov:2000qh}, and then run it
at two-loop order numerically to the indicated scale
$\mu_t$ \cite{Carena:1999py}.

\subsection{Real radiation contribution}
\label{subsec2real}
As already discussed in subsection \ref{subsecpowheg},
in the \POWHEGBOX framework the real emission
channels are assigned to the $R^{\text{div}}$ or $R^{\text{reg}}$ groups, depending
on whether they feature soft/collinear-divergent behavior or not.

In  double Higgs production in gluon fusion, the   $gg \to h h g$ and
$gq \to h h q$ channels belong to $R^{\text{div}}$,
while the  $q\bar{q} \to hhg$ channel belongs to  $R^{\text{reg}}$.

The implementation of these channels has been achieved using the {\tt MadLoop}
matrix element generator \cite{Hirschi:2011pa}. The code generated
by {\tt MadLoop} is interfaced
directly with the \POWHEGBOX. The SM model file shipped with {\tt MadLoop} has
been modified in such a way to include an additional parameter that allows for
a rescaling of the Higgs trilinear coupling.

To study the uncertainties in the matching of the NLO calculation  with parton
showers we use the possibility given by the \POWHEGBOX
framework to split the contribution of the  processes in $R^{\text{div}}$ into
a singular and a finite contribution, respectively $R^s$ and $R^f$. The events
that contribute to the  $R^f$ term are called ``remnant events''.

The separation of  $R^{\text{div}}$ in $R^s$ and $R^f$ 
is achieved dynamically by using a damping factor, $D_h$, via
\begin{align}
  R^s = D_h~R_{\text{div}} ~,\phantom{aaaaaaaaaaa}
  R^f = \left(1-D_h\right)~R_{\text{div}}\,.
\end{align}
with
\begin{align}
  D_h = \frac{h^2}{h^2+(p_{\bot}^{HH})^2}\,,
  \label{eq:resframework:dh}
\end{align}
where $p_{\bot}^{HH}$ is the transverse momentum of the two-Higgs system,
and the default value in \POWHEG for $h$ is $h= \infty$.

Once this separation has been performed, another freedom present
in the \POWHEGBOX framework is the choice of the shower scale for the
remnant events (we recall that varying this scale is a higher order effect).
By default this scale is set to the $p_T$ of the radiated parton. However, it
has been found during the study of single Higgs production in gluon
fusion~\cite{Bagnaschi:2015qta,Bagnaschi:2015bop},
that such a choice yields, at large $p_T$, harder tails for the NLO
system with respect to the fixed order result.  To
recover the fixed order behavior, it is possible to choose lower
scales, in order to limit the phase space available for further
emissions by the shower.

We conclude this section by comparing the performance of our MC code
with that of the code \ggHH presented in  Ref.~\cite{Heinrich:2019bkc}.
In the latter the virtual two-loop corrections are implemented via several
numerical grids \cite{GitHub} in the Mandelstam variables $\hat{s}$ and
$\hat{t}$ for fixed
values of the top and Higgs mass and $\alpha_s$. An interpolation framework is
also provided in order to produce the virtual two-loop amplitude at any
point in the phase space. This procedure is faster than our approach to compute
the corrections at any point in the phase space via our analytic result.
However, it lacks flexibility in the input parameters and, in some regions
of the phase space, sometimes the interpolated result is not very accurate \cite{Alasfar:2023xpc}.

On the contrary, for what concern the real emission contributions our MC code
is faster than \ggHH. This is  due to the different way these contributions
are computed. We use the {\tt MadLoop} code to compute the real radiation
matrix elements while in \ggHH  the same contribution is computed using the
{\tt GoSam} code \cite{GoSam:2014iqq,Cullen:2011ac}.

The net result of these two competing factors is that our MC has an average
timing for evaluating one phase-space point slightly shorter than the
corresponding timing in \ggHH.

\section{Results}
\label{sec:res}
In this section, we present our numerical results for a center-of-mass
energy $\sqrt{s}=13.6\text{ TeV}$. The values of the input parameters
are chosen according to the latest recommendation of the LHC Higgs Working
Group (LHCHWG):
\begin{align}
  &m_t^{\textrm{OS}}=\,172.5~\mathrm{GeV}, \quad \quad m_W=80.385\,\mathrm{GeV},
  \quad \quad m_H=125\,\mathrm{GeV}, \nn \\
  &G_{\mu}=1.1663787\times10^{-5}\,{\mathrm{ GeV}^{-2}}\,.
\end{align}
For our studies we adopt the {\tt NNPDF31\_nlo\_as\_0118} \cite{NNPDF:2017mvq}
parton distribution functions (PDF) in a five flavour scheme  as reference
PDF set for the NLO calculation. Correspondingly, the LO results are  obtained
using the same set extracted at LO.
The value of strong coupling constant is set to be the same as the one
used in the PDF, i.e. $\alpha_s(M_Z) = 0.118$.


\subsection{Inclusive Cross Section}
In Table \ref{tab:totxs}, we show the total cross section at 13.6 TeV
at LO and NLO for several values of $\tril$ adopting different top-quark-mass
renormalization schemes, i.e.~OS
and $\MSbar$ with different scale choices $\mu_t$.  The  values
of the renormalization and factorization scales are fixed to be
$\mu_C=M_{HH}/2$ and the scale uncertainty is estimated from the envelope
of a 7-point rescaling of $\mu_C$ according to
$(\mu_R/\mu_C,\mu_F/\mu_C)=(1,1),(1,\frac{1}{2}),(1,2),(\frac{1}{2},
\frac{1}{2}),(\frac{1}{2},1),(2,1),(2,2)$.

\begin{table}
  \centering
  \begin{tabular}{c|c|c|c|c|c}
$\tril/\trilsm$ &      Top-mass scheme & LO [fb] & $\sigma_{LO}/\sigma_{LO}^{OS}$ & NLO [fb] & $\sigma_{NLO}/\sigma_{NLO}^{OS}$ \\
\hline
-0.6 & On-Shell & $55.76_{ -21.0\% }^{ 28.7\% }$ & - & $100.77_{ -13.7\% }^{ 15.8\% }$ & - \\
\hline
-0.6 & $\MSbar,\mu_t=M_{HH}/4$ & $53.52_{ -20.9\% }^{ 28.6\% }$ & 0.96 & $98.52_{ -13.9\% }^{ 16.2\% }$ & 0.98 \\
-0.6 & $\MSbar,\mu_t=M_{HH}/2$ & $54.48_{ -20.8\% }^{ 28.5\% }$ & 0.98 & $99.14_{ -13.8\% }^{ 16.1\% }$ & 0.98 \\
-0.6 & $\MSbar,\mu_t=M_{HH}$ & $55.21_{ -20.7\% }^{ 28.3\% }$ & 0.99 & $99.63_{ -13.7\% }^{ 16.1\% }$ & 0.99 \\
-0.6 & $\MSbar,\mu_t=m_t^{\MSbar}(m_t^{\MSbar})$ & $56.68_{ -20.9\% }^{ 28.5\% }$ & 1.02 & $101.07_{ -13.5\% }^{ 15.5\% }$ & 1.00 \\
\hline
0 & On-Shell & $38.50_{ -21.1\% }^{ 28.9\% }$ & - & $68.38_{ -13.4\% }^{ 15.1\% }$ & - \\
\hline
0 & $\MSbar,\mu_t=M_{HH}/4$ & $36.65_{ -21.0\% }^{ 28.8\% }$ & 0.95 & $66.52_{ -13.7\% }^{ 15.8\% }$ & 0.97 \\
0 & $\MSbar,\mu_t=M_{HH}/2$ & $36.75_{ -20.9\% }^{ 28.7\% }$ & 0.95 & $66.40_{ -13.7\% }^{ 15.8\% }$ & 0.97 \\
0 & $\MSbar,\mu_t=M_{HH}$ & $36.70_{ -20.8\% }^{ 28.5\% }$ & 0.95 & $66.30_{ -13.7\% }^{ 15.9\% }$ & 0.97 \\
0 & $\MSbar,\mu_t=m_t^{\MSbar}(m_t^{\MSbar})$ & $38.48_{ -21.0\% }^{ 28.7\% }$ & 1.00 & $67.87_{ -13.3\% }^{ 15.1\% }$ & 0.99 \\
\hline
1 & On-Shell & $18.22_{ -21.3\% }^{ 29.5\% }$ & - & $30.93_{ -12.7\% }^{ 13.7\% }$ & - \\
\hline
1 & $\MSbar,\mu_t=M_{HH}/4$ & $16.94_{ -21.3\% }^{ 29.3\% }$ & 0.93 & $29.68_{ -13.2\% }^{ 14.7\% }$ & 0.96 \\
1 & $\MSbar,\mu_t=M_{HH}/2$ & $16.22_{ -21.2\% }^{ 29.1\% }$ & 0.89 & $28.90_{ -13.5\% }^{ 15.2\% }$ & 0.93 \\
1 & $\MSbar,\mu_t=M_{HH}$ & $15.48_{ -21.1\% }^{ 29.0\% }$ & 0.85 & $28.27_{ -13.9\% }^{ 16.1\% }$ & 0.91 \\
1 & $\MSbar,\mu_t=m_t^{\MSbar}(m_t^{\MSbar})$ & $17.30_{ -21.2\% }^{ 29.2\% }$ & 0.95 & $29.78_{ -13.0\% }^{ 14.3\% }$ & 0.96 \\
\hline
2.4 & On-Shell & $7.68_{ -21.3\% }^{ 29.3\% }$ & - & $13.41_{ -13.1\% }^{ 14.8\% }$ & - \\
\hline
2.4 & $\MSbar,\mu_t=M_{HH}/4$ & $7.01_{ -21.2\% }^{ 29.1\% }$ & 0.91 & $12.84_{ -13.8\% }^{ 16.3\% }$ & 0.96 \\
2.4 & $\MSbar,\mu_t=M_{HH}/2$ & $6.43_{ -21.1\% }^{ 28.9\% }$ & 0.84 & $12.42_{ -14.6\% }^{ 17.7\% }$ & 0.93 \\
2.4 & $\MSbar,\mu_t=M_{HH}$ & $6.00_{ -21.0\% }^{ 28.7\% }$ & 0.78 & $12.07_{ -15.1\% }^{ 18.5\% }$ & 0.90 \\
2.4 & $\MSbar,\mu_t=m_t^{\MSbar}(m_t^{\MSbar})$ & $6.92_{ -21.1\% }^{ 29.1\% }$ & 0.90 & $12.81_{ -14.1\% }^{ 16.3\% }$ & 0.96 \\
\hline
6.6 & On-Shell & $101.00_{ -20.2\% }^{ 27.4\% }$ & - & $203.91_{ -15.2\% }^{ 19.0\% }$ & - \\
\hline
6.6 & $\MSbar,\mu_t=M_{HH}/4$ & $100.81_{ -20.2\% }^{ 27.4\% }$ & 1.00 & $203.90_{ -15.2\% }^{ 19.1\% }$ & 1.00 \\
6.6 & $\MSbar,\mu_t=M_{HH}/2$ & $109.75_{ -20.2\% }^{ 27.4\% }$ & 1.09 & $213.28_{ -14.7\% }^{ 18.3\% }$ & 1.05 \\
6.6 & $\MSbar,\mu_t=M_{HH}$ & $119.17_{ -20.2\% }^{ 27.4\% }$ & 1.18 & $221.06_{ -14.1\% }^{ 17.3\% }$ & 1.08 \\
6.6 & $\MSbar,\mu_t=m_t^{\MSbar}(m_t^{\MSbar})$ & $110.59_{ -20.2\% }^{ 27.4\% }$ & 1.09 & $214.57_{ -14.7\% }^{ 18.3\% }$ & 1.05 \\
\end{tabular}

  \caption{Total cross section for the Higgs boson pair production at
   $\sqrt{s} =13.6$ TeV for several values of $\tril$.  The LO and NLO
   results are
   shown using different
   top-quark-mass renormalization schemes. The central value of the
   renormalization and factorization scales is fixed to be
   $\mu_R=\mu_F=M_{HH}/2$. Scale uncertainties are taken from a
   7-point scale variation.}
 \label{tab:totxs}
\end{table}

We find that the NLO corrections are large for each choice of the
top-mass renormalization scheme.  Moreover, the relative
 size of the scale uncertainties is essentially the same regardless of
the top-mass renormalization scheme employed.  We note that going from LO to
NLO the relative size of the scale uncertainties is reduced by a
factor of $\sim 40\%$.

\begin{figure}[t]
  \centering
  \includegraphics[scale=0.5]{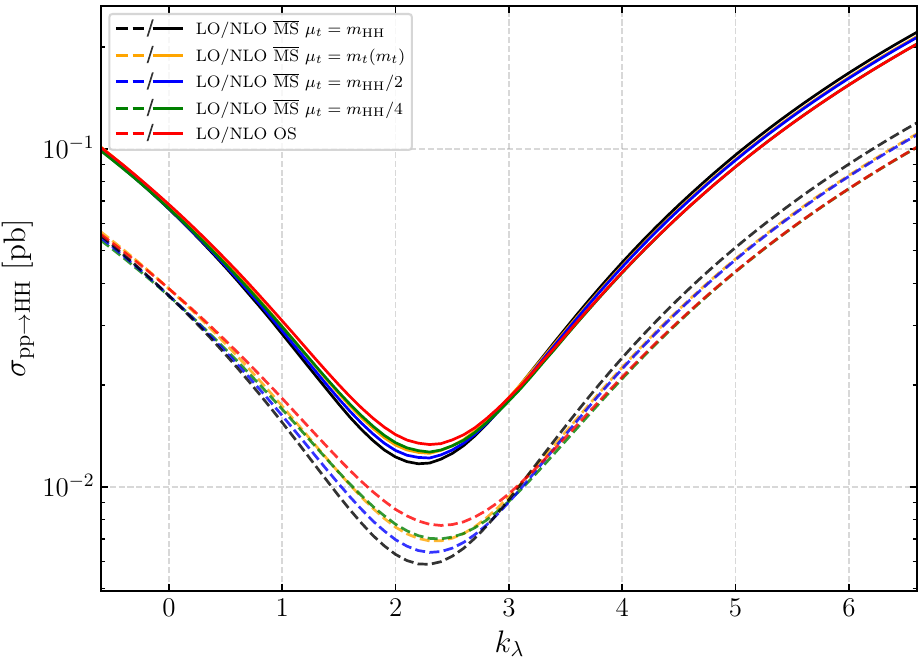}~\includegraphics[scale=0.5]{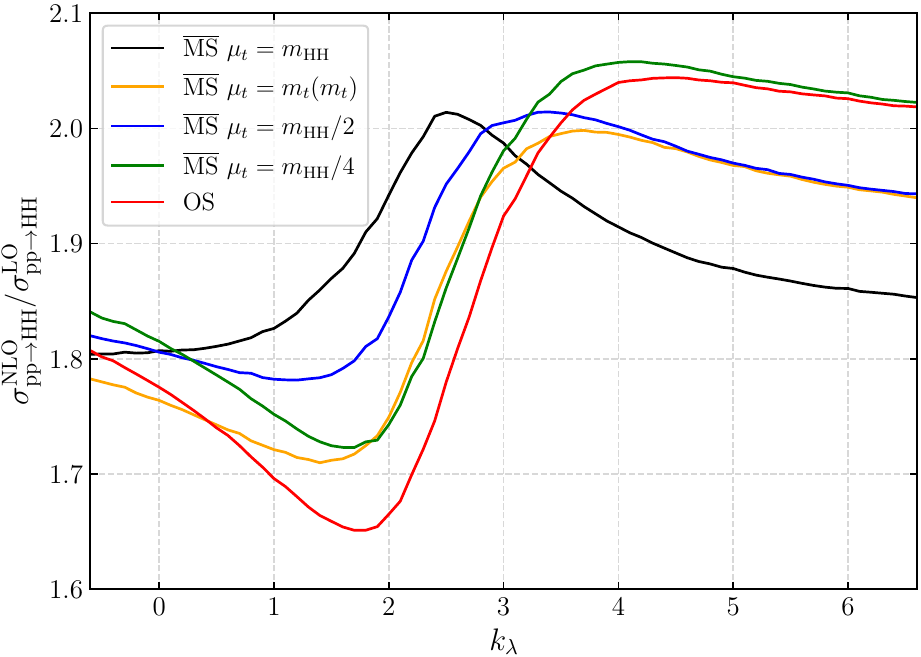}
  \caption{Left: the total inclusive cross sections at $\sqrt{s}= 13.6$ TeV
    for different choices of the top mass renormalization scheme, at LO
    (dashed) and NLO (solid), as a function $\ktre$.
    Right: the corresponding $K$-factors.}
\label{fig:Kfac}
\end{figure}

As illustrated in the left panel of fig.~\ref{fig:Kfac}, where we show
the inclusive cross section at LO and NLO as a function of the Higgs
trilinear coupling for different top-mass scheme, the OS scheme leads
to the largest value of the total cross section both at LO and NLO up
to $\ktre \sim 3$. In the same range of $\ktre$ the $\MSbar$ scheme
for $\mu_t=M_{HH}$ gives the smallest cross section while as $\ktre$
increases tends to give the largest one. The maximum difference
between the schemes is obtained around the minimum of the cross-section,
which corresponds to the maximal destructive interference between the
``signal'' and ``background'' diagrams. 
At LO, the maximum difference between the schemes amounts to about 20\%
(for $\ktre = 2.4$), while it decreases to 10\%
at NLO (again for $\ktre = 2.4$).  We notice that the exact $\ktre$ value that
gives the minimum
of the cross section actually depends upon the top-mass scheme. As
expected, going from LO to NLO the various minima get closer.

In the right panel of fig.~\ref{fig:Kfac},  we show  the  $K$-factors,
$K=\sigma_{NLO}/\sigma_{LO}$, as a function of the Higgs trilinear coupling for
different top-mass scheme. We recall that going from a LO result to a NLO one
the scheme dependence is expected to decrease. Then, schemes where the LO
prediction are smaller show the largest  $K$-factors.

Let us now comment on our findings for the inclusive cross as a function
od $\tril$ compared to the same study carried out with the \ggHH MC code
of Ref.~\cite{Heinrich:2019bkc} and with the results presented in
Ref.~\cite {Baglio:2020wgt}.

The comparison with \ggHH was done adopting for the top and Higgs
masses and $\as$ the values chosen in \ggHH. While we found excellent
agreement within the numerical errors of the MCs for the SM, during
the course of this study, we have identified a discrepancy with the
existing calculation of Ref.~\cite{Heinrich:2019bkc} when varying
$\kappa_{\lambda}$ away from its SM value. We have traced this
discrepancy to the two-loop virtual contributions. We contacted the
authors of ref.~\cite{Heinrich:2019bkc} who, using our results, found
indeed an issue in their two-loop virtual contributions for values of
the trilinear coupling different from the SM one. The authors of
ref.~\cite{Heinrich:2019bkc} provided us with a fixed version of their
calculation. Using these new results, we found now agreement between
the two codes.

For the comparison with the results of Ref.~\cite {Baglio:2020wgt} we adopted
the same set of PDF employed in that work.
For the SM cross section  we are in agreement with the results of
Ref.~\cite {Baglio:2020wgt} at  various center-of- mass energies including
their estimate of the  scale uncertainty. Instead,
concerning the dependence of
the cross section upon $\ktre$, we find an agreement with the results of
Ref.~\cite {Baglio:2020wgt} at the level of per-mille
for negative values of $\ktre$ and for $\ktre =1$. For larger positive values
of $\ktre$ the agreement starts to deteriorate. At the minimum of the
cross section, $\ktre \sim 2.4$, we find the maximal discrepancy between our
and their evaluation of  the inclusive cross section with a difference of
several per-cent. This discrepancy  is difficult to trace. On one side, the
agreement  with  \ggHH, after the fixing by the authors, and
  the agreement with Ref.~\cite {Baglio:2020wgt}
for $\ktre \le 1$ gives us confidence in our MC code. On the other
side, the fact that the maximum difference is obtain at the minimum of
the cross section, where the destructive interference between ``signal'' and
``background'' contributions is more pronounced, let us  suspect that the
way the inclusive cross section is computed in Ref.~\cite {Baglio:2020wgt}
from finite size bins, could be not sufficiently accurate in regions of the
parameter space where there are strong cancellations. This is also suggested
by the fact that we find that the discrepancy with Ref.~\cite {Baglio:2020wgt}
decreases for large positive value of $\ktre$.



\subsection{Differential Distributions}

We now present the results for a selection of differential observables.
As in the previous subsection we focus on the dependence of the  observables
upon the choice of the top-quark-mass renormalization schemes.
First, we consider  the SM case, $\ktre =1$,  then we repeat the analyis
for several values of $\ktre$ in order to  investigate the interplay between
the renormalization scheme choice and the value of the Higgs trilinear
coupling.

\subsubsection{Renormalization scheme dependence in the SM}

\begin{figure}
  \includegraphics[width=0.5\textwidth]{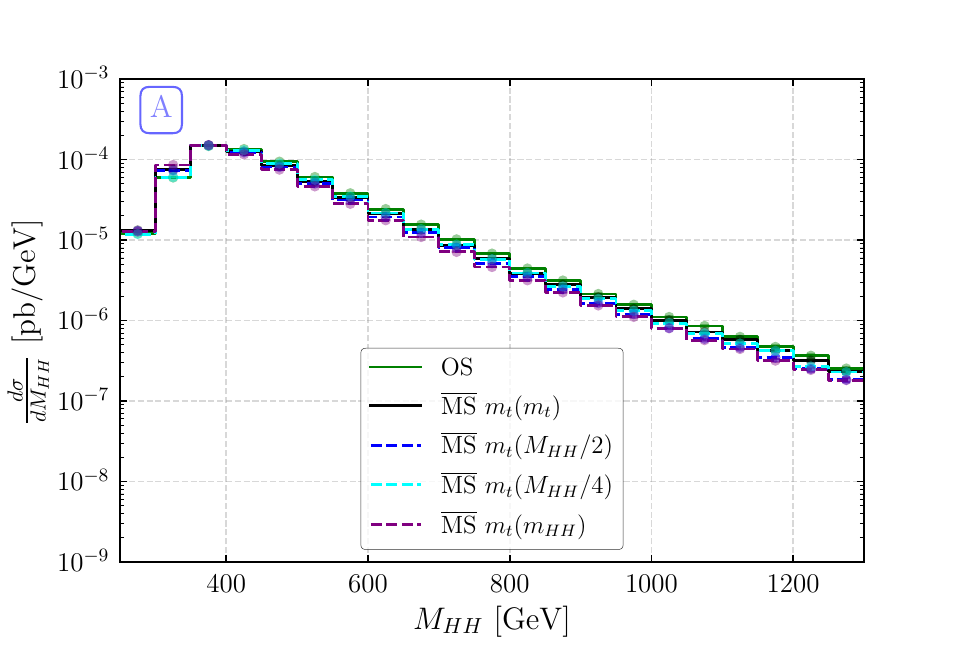}~\includegraphics[width=0.5\textwidth]{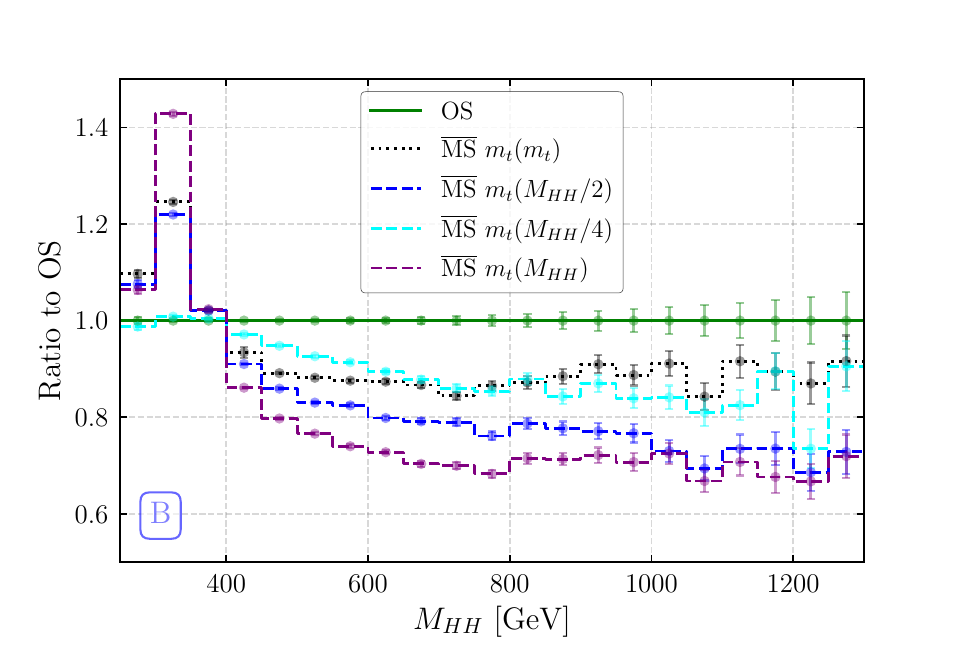}\\
  \includegraphics[width=0.5\textwidth]{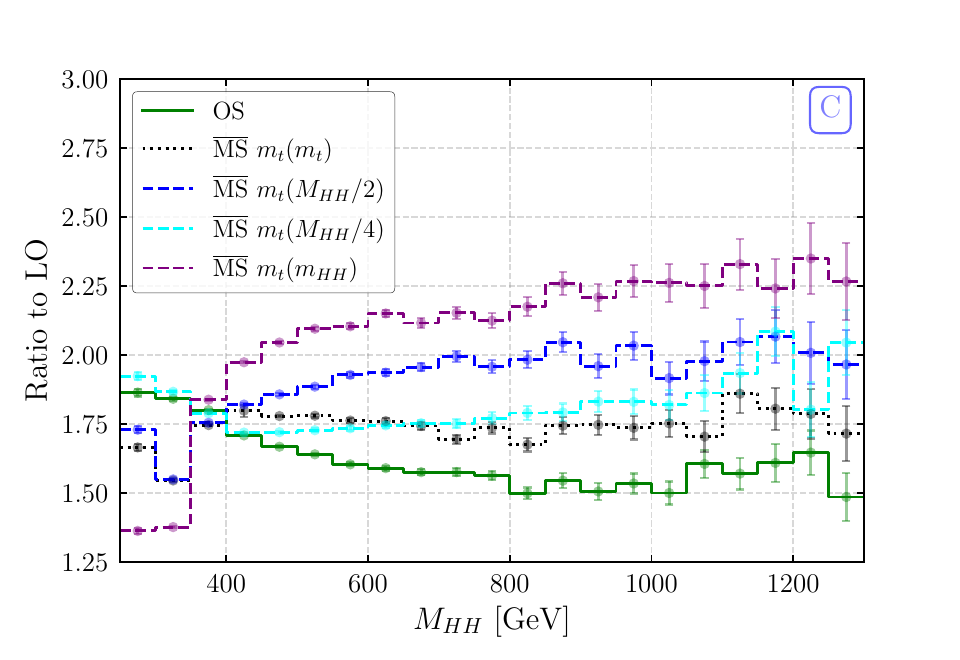}~\includegraphics[width=0.5\textwidth]{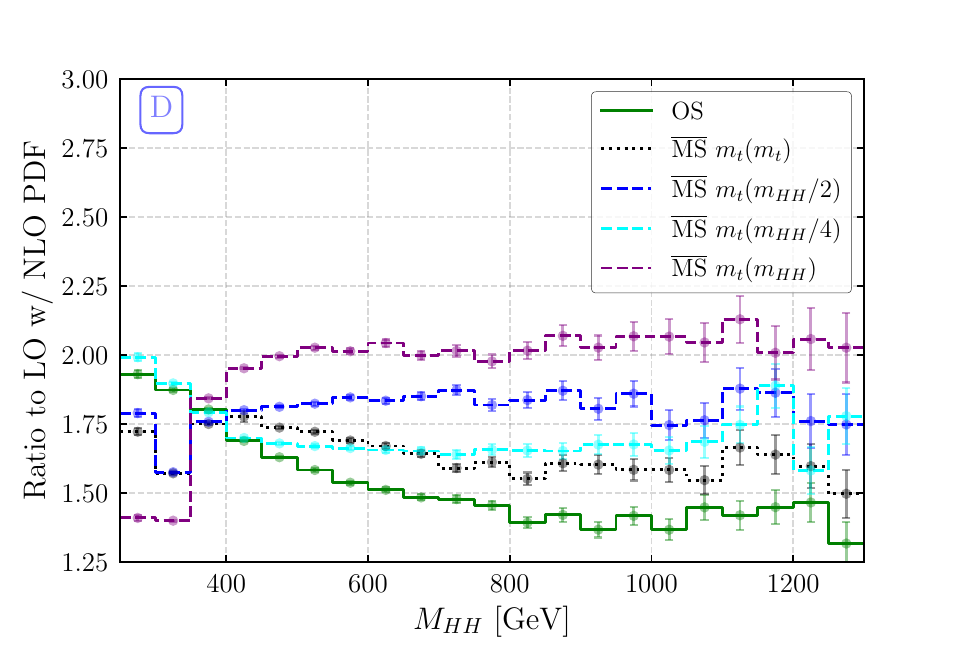}
  \caption{The invariant mass distribution of the two-Higgs system for
    different choices of the top-mass renormalization scheme:  (\textbf{A})
    absolute distributions at NLO+PS; (\textbf{B}) ratio between the
    $\MSbar$ predictions  and the  OS one;  (\textbf{C})
    ratio between the distributions computed
    at NLO+PS and their LO counterpart ($K$-factors);
    (\textbf{D}) same as \textbf{C} but with the LO distributions
    computed with NLO PDFs.}
 \label{fig:mhh}
 \end{figure}

In fig.~\ref{fig:mhh} we show the $M_{HH}$ distribution adopting
different renormalization schemes for the top mass. This and the
following figures are obtained at the NLO+PS level using our \POWHEG
MC code\footnote{The {\tt hdamp} parameter in \POWHEG is set to its
  default value, {\tt hdamp}=$\infty$ ($D_h=1$).}  in conjunction with
the {\tt Pythia 8} shower \cite{Sjostrand:2007gs,Sjostrand:2014zea}.
In the top left figure, \textbf{A}, we present the absolute distribution that
shows the peak at the opening of the $2 \,\mt$ threshold. Although not
clearly visible in the figure, the position of the peak actually
depends on the top mass scheme.  In fig.~\ref{fig:mhh}~\textbf{B}, the ratio
between the $\MSbar$ predictions and the OS one is shown.  We notice
that for large values of $M_{HH}, \,(M_{HH} \geq 600$ GeV) this ratio
is approximately constant with the $\mt(M_{HH})$ scheme giving the
smallest cross section. The $\MSbar \, \mt(M_{HH}/4)$ values are up to
$M_{HH} \sim 700$ GeV the closest to the OS ones.  In this $M_{HH}$
range the $\mt(M_{HH}/4)$ scheme is the one where the running of the
top mass is the least. Below $M_{HH} \sim 400$ GeV the other $\MSbar$
predictions show significant deviations from the OS one especially in
the region around the opening of the $2 \,\mt$ threshold.  As already
said, the opening of the $2 \,\mt$ threshold occurs at different
values of $M_{HH}$ in the various top mass schemes and therefore the
ratio shown is influenced by the position of the peak.  The bottom
part of fig.~\ref{fig:mhh} shows the $K$-factor in the various
schemes. In fig.~\ref{fig:mhh}~\textbf{C} the LO result is evaluated using the
LO PDF while in the NLO result the NLO PDF is used. In
fig.~\ref{fig:mhh}~\textbf{D}
instead, the NLO PDF is used both at LO and NLO. As already said going from a
LO result to a NLO one the scheme dependence is expected to decrease.
Then if an $\MSbar$ scheme has a LO prediction smaller than the corresponding OS
one its $K$-factor is expected to be larger than the OS $K$-factor and
vice versa. Taking into account the information shown in
fig.~\ref{fig:mhh}~\textbf{B}, one sees that the behavior of the $K$-factors
in the bottom left figure shows exactly this feature. In
fig.~\ref{fig:mhh}~\textbf{D}, the effects due to the variation of the PDFs
are eliminated so
that the pure scheme-dependent effects are more manifest.  The effect
induced by the variation of the PDF is quite mild as can be seen
comparing figs.~\ref{fig:mhh}~\textbf{C} with \ref{fig:mhh}~\textbf{D}.

\begin{figure}
  \includegraphics[width=0.5\textwidth]{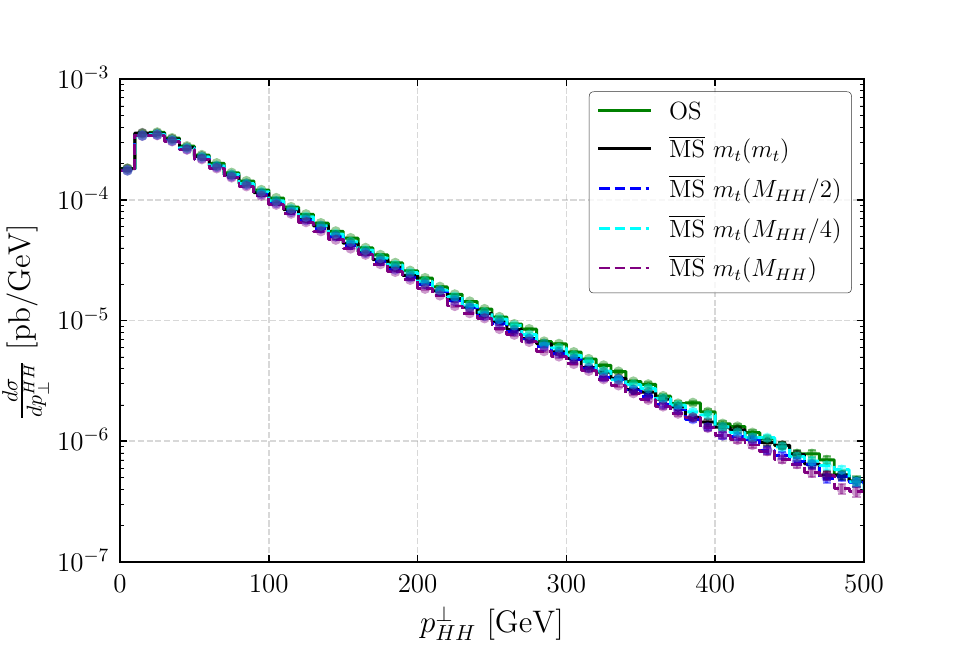}~\includegraphics[width=0.5\textwidth]{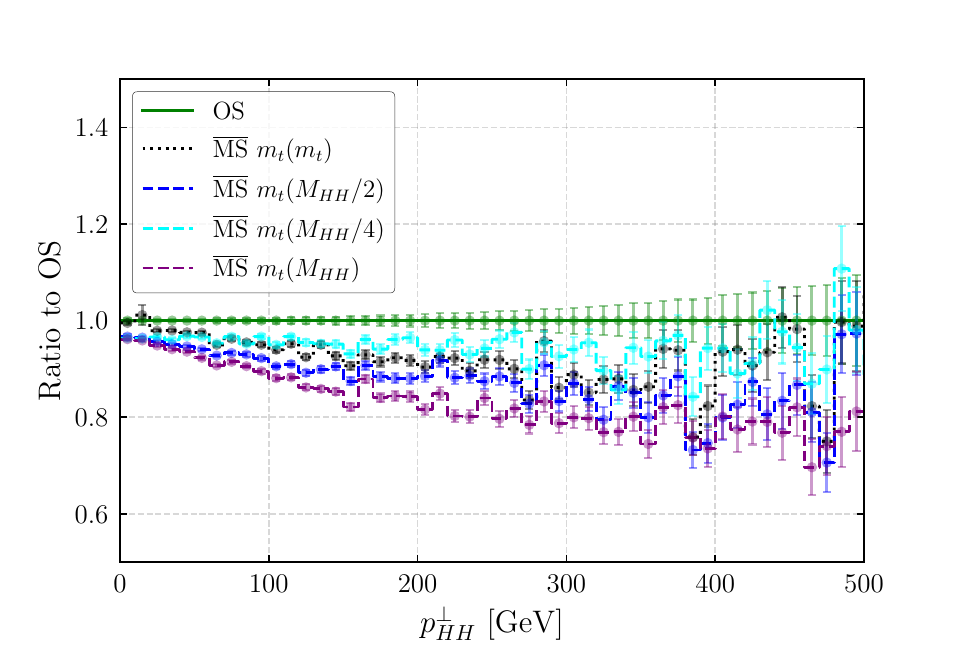}\\
  \caption{The transverse momentum distribution of the two-Higgs system for
    different choices of the top-mass renormalization scheme: (left)
    absolute distributions at NLO-PS; (right) ratio between the
    $\MSbar$ predictions  and the  OS one.}
  \label{fig:ptHH}
\end{figure}

The $M_{HH}$ observable is quite insensitive to shower effects. To appreciate
the latter we consider two observable which are sensitive to the recoil against
jet activity, namely the transverse momentum of the two Higgs system,
$p^\perp_{HH}$, and the  transverse momentum of the leading Higgs, $p^\perp_{H_{1}}$,
that is identified as the final state boson with the largest transverse
momentum.

In fig.~\ref{fig:ptHH} we show the transverse momentum distribution of the two
Higgs system  for different top mass renormalization schemes.
This observable is sensitive to soft gluon radiation. In the left figure
the absolute distributions at NLO+PS are presented which show the suppression
in the region $p^\perp_{HH} \to 0$ where the fixed order NLO results become
unreliable. The right plot shows the ratio between the $\MSbar$ predictions
for the $p^\perp_{HH}$  distribution and the OS one. We notice that
the $\MSbar$ results are always smaller than the OS one. However, in the
small $p^\perp_{HH}$  region they are all quite close to the OS result while,
as the $p^\perp_{HH}$ increases the $\MSbar$ results tend to be more spread
among themselves and with respect to the OS one. The small scheme dependence
as $p^\perp_{HH} \to 0$ is likely related to the shower, that in this region
has a relevant effect.
\begin{figure}
  \includegraphics[width=0.5\textwidth]{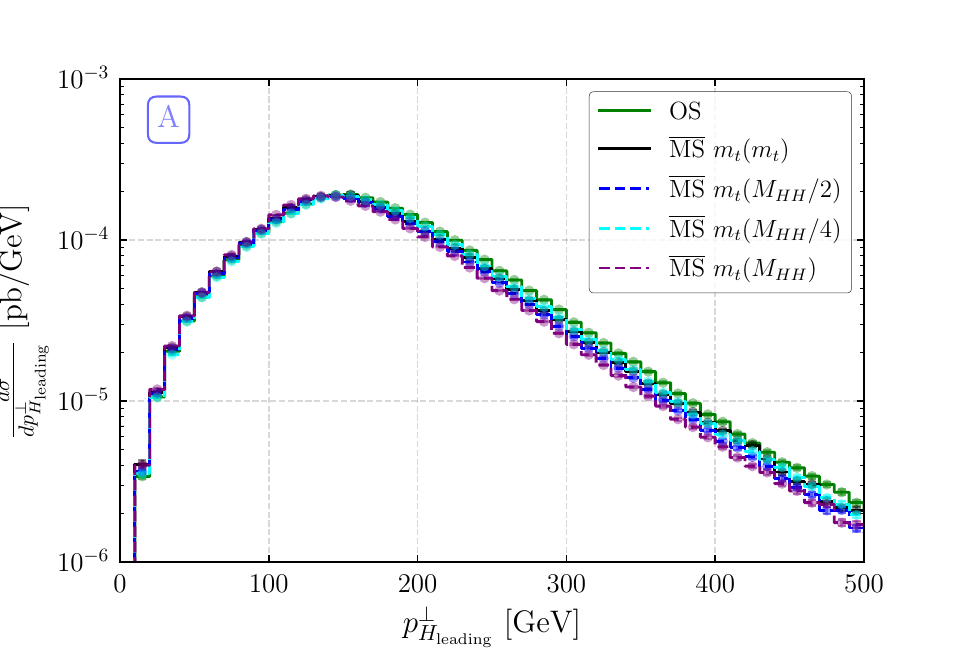}~\includegraphics[width=0.5\textwidth]{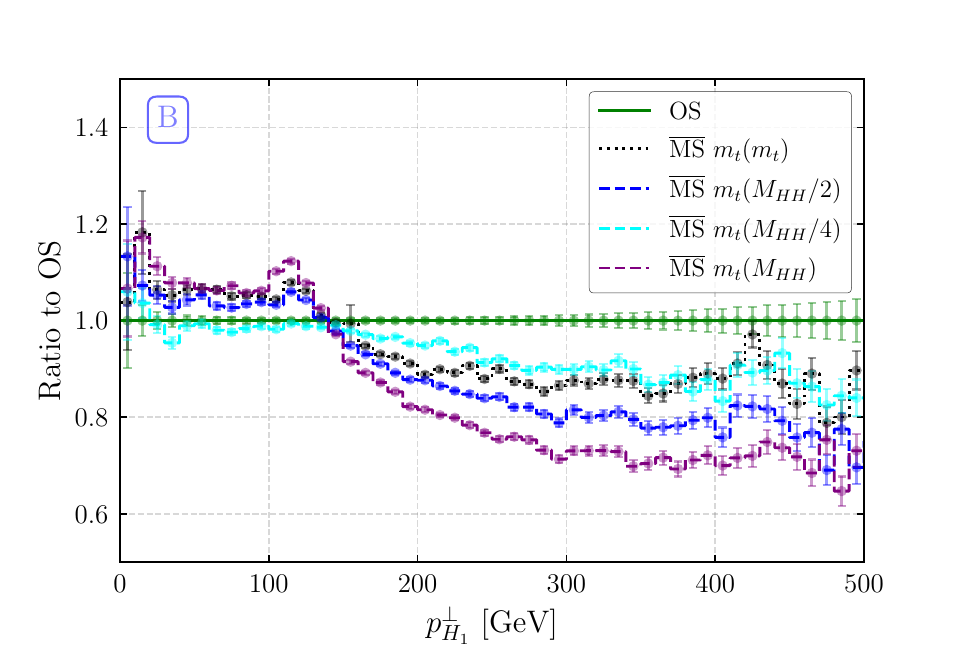}\\
  \includegraphics[width=0.5\textwidth]{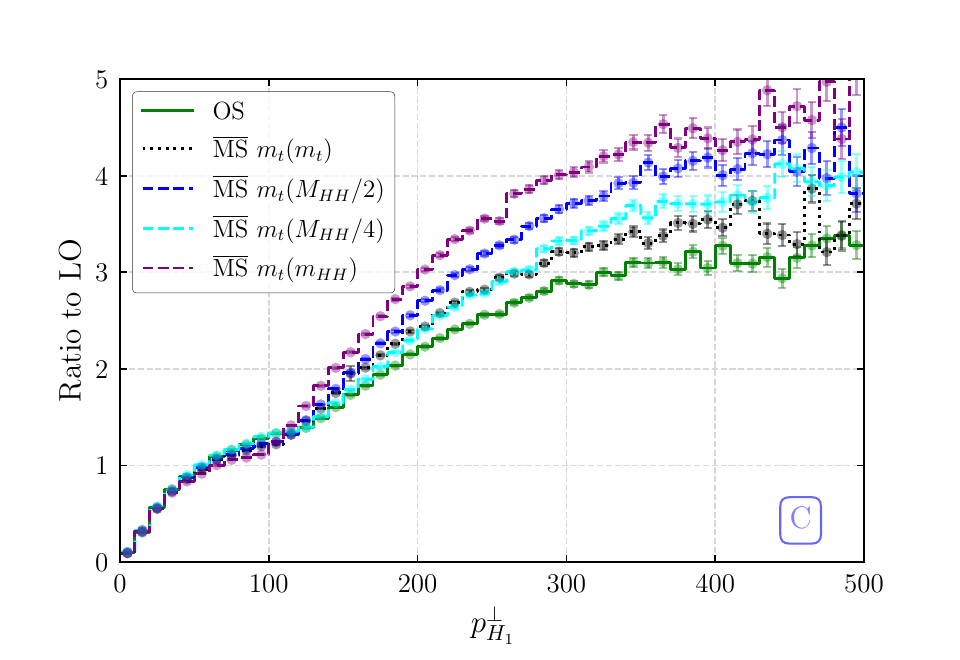}~\includegraphics[width=0.5\textwidth]{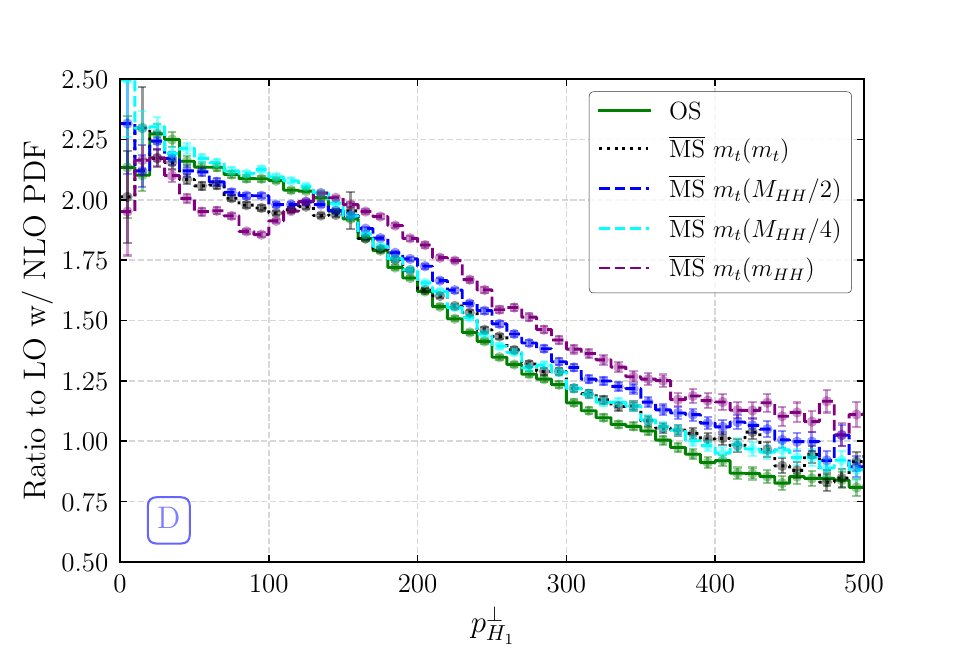}
  \caption{The transverse momentum of the leading Higgs for
    different choices of the top-mass renormalization scheme: (\textbf{A})
    absolute distributions at NLO+PS; (\textbf{B}) ratio between the
    $\MSbar$ predictions  and the  OS one; (\textbf{C})
    ratio between the distributions computed
    at NLO+PS and their LO counterpart ($K$-factors);
    ( \textbf{D}) same as bottom left, but with the LO distributions
    computed with NLO PDFs.}
  \label{fig:ptH1}
\end{figure}

Fig.~\ref{fig:ptH1} shows the same plots of fig.~\ref{fig:mhh} for the
transverse momentum of the leading Higgs instead of $M_{HH}$.
Figs.~\ref{fig:ptH1}~\textbf{B} and \ref{fig:ptH1}~\textbf{C}
convey the same information with respect to the scheme
dependence as the corresponding plots in fig.~\ref{fig:mhh}.
In fig.~\ref{fig:ptH1}~\textbf{C} all the $K$-factors
are quite close up to $p^\perp_{H_{1}} \sim 100$ GeV
showing that the $p^\perp_{H_{1}} $ predictions in this region are little
scheme-dependent. For higher values of  $p^\perp_{H_{1}} $ the scheme dependence
in more pronounced and the $K$-factors can be very large. However,
fig.~\ref{fig:ptH1}~\textbf{D}, which is the same as plot \textbf{C} but with
LO and NLO
distributions computed with NLO PDFs, shows that the choice of the PDF
plays an important role and, once the same PDF is used both at LO and NLO,
the $K$-factors become smaller and the scheme dependence is approximately
the same for all values of $p^\perp_{H_{1}} $.

\subsubsection{Interplay between a modified trilinear coupling and the
  renormalization scheme}
In this subsection the same analyses presented in the previous subsection
are repeated for different values of the Higgs trilinear coupling. For shortness
we concentrate on the ratio between the $\MSbar$ predictions and the OS one at
NLO+PS.

\begin{figure}
  \includegraphics[width=0.5\textwidth]{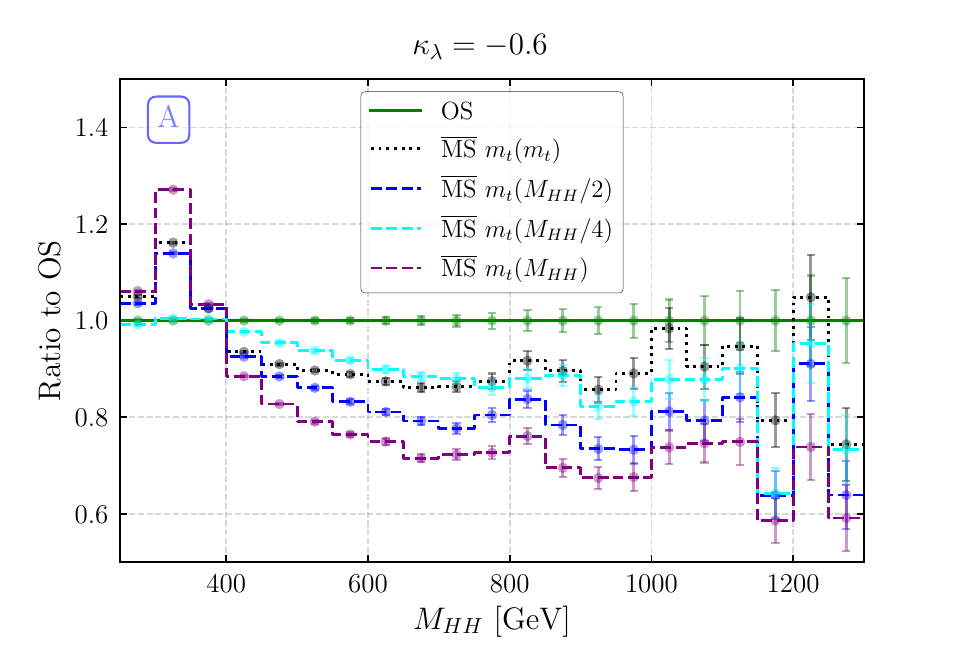}~\includegraphics[width=0.5\textwidth]{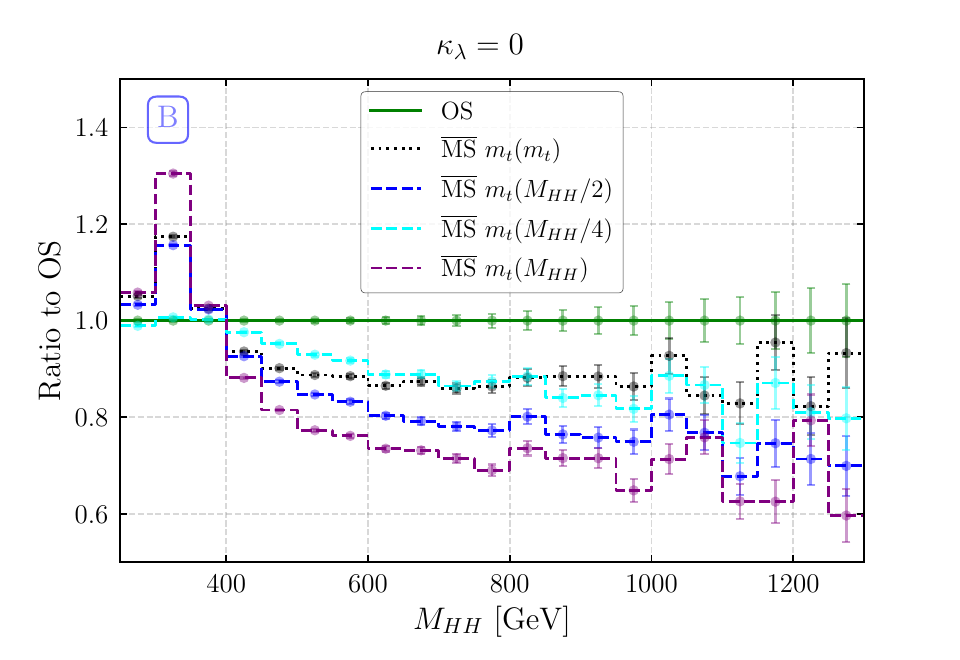}\\
  \includegraphics[width=0.5\textwidth]{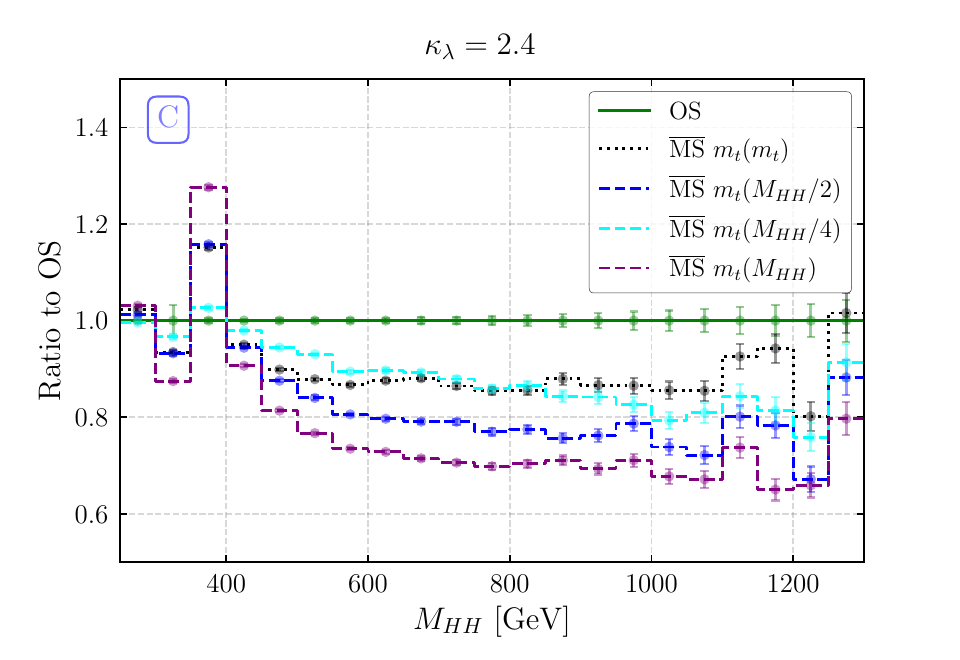}~\includegraphics[width=0.5\textwidth]{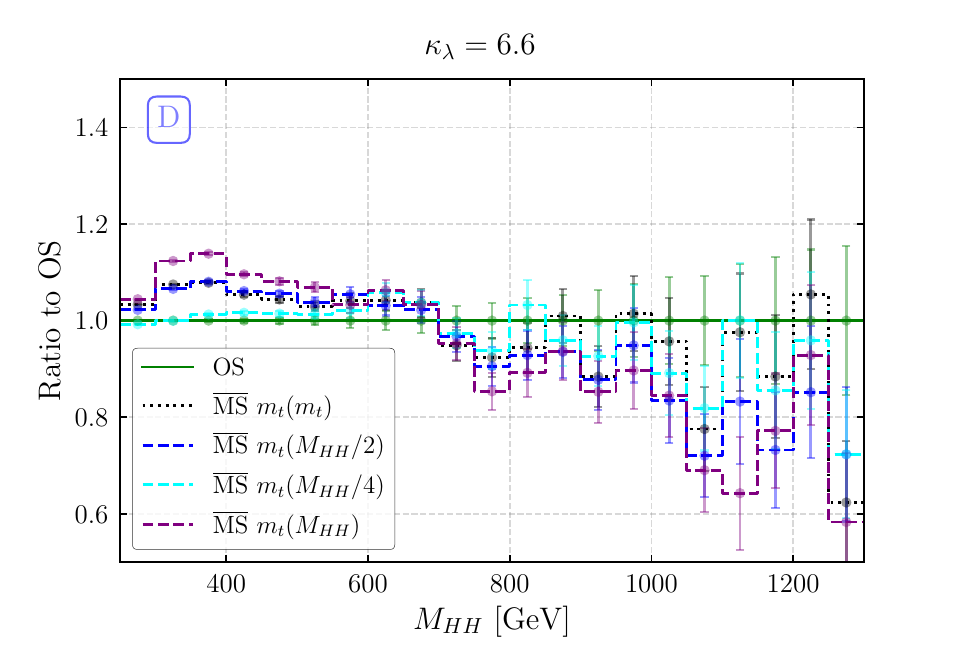}
  \caption{The  invariant mass distribution of the two-Higgs system
    for several values of $\ktre$ and different choices of the top-mass
    renormalization scheme: ratio between the
    $\MSbar$ predictions  and the  OS one for $\ktre = -0.6$ (\textbf{A});\,
    $\ktre = 0$ (\textbf{B}), $\ktre = 2.4$ (\textbf{C}),
    $\ktre = 6.6$ (\textbf{D}).}
  \label{fig:mhh-klambda}
\end{figure}

In fig.~\ref{fig:mhh-klambda} the invariant mass distribution of the two Higgs
system is considered. The four plots in the figure have to be compared with
fig.~\ref{fig:mhh}~\textbf{A} that correponds to the case
$\ktre = 1$. The figure shows that the
cases $\ktre = -0.6$ (\textbf{A}) and $\ktre =0$ (\textbf{B}) are very similar
to $\ktre =1$. The case $\ktre =0$, where there is no ``signal'' contribution,
indicates that the scheme dependence of the ``signal'' part is much milder
than that of the ``background'' one.
Instead, for $\ktre= 2.4$ (\textbf{C}) i.e. the $\tril$ value where  the
negative
interference between the ``signal'' and ``background'' diagrams in the OS
scheme is maximal,  one sees that the scheme dependence is more pronounced
in the region around the $2\, \mt$ threshold with repect to the SM case.
Finally, the case $\ktre = 6.6$ (\textbf{D}) shows a much
milder scheme dependence, as expected, because in this case the ``signal''
contribution is very amplified.

\begin{figure}
  \includegraphics[width=0.5\textwidth]{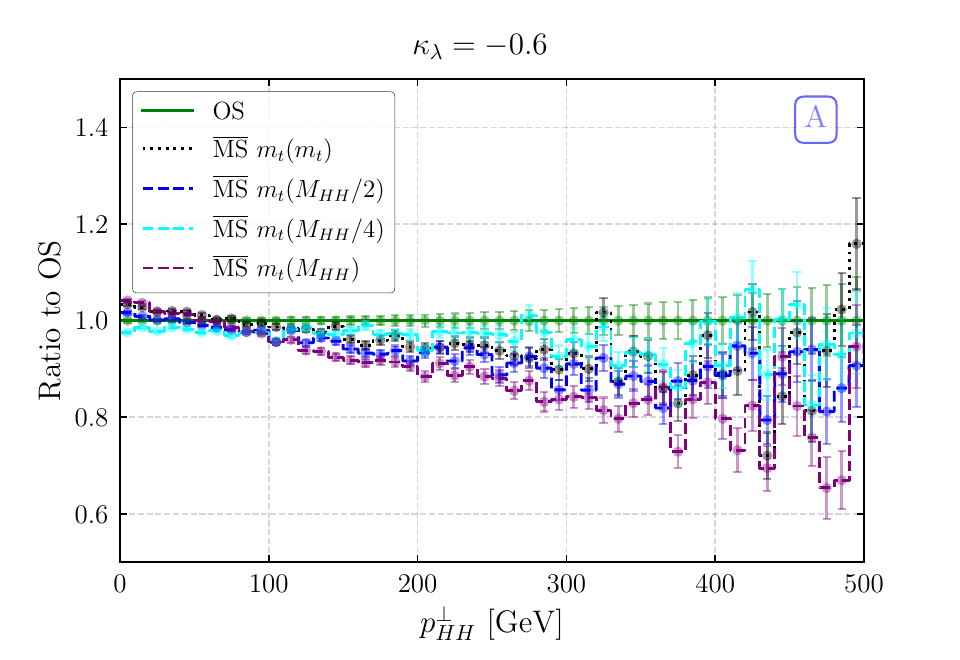}~\includegraphics[width=0.5\textwidth]{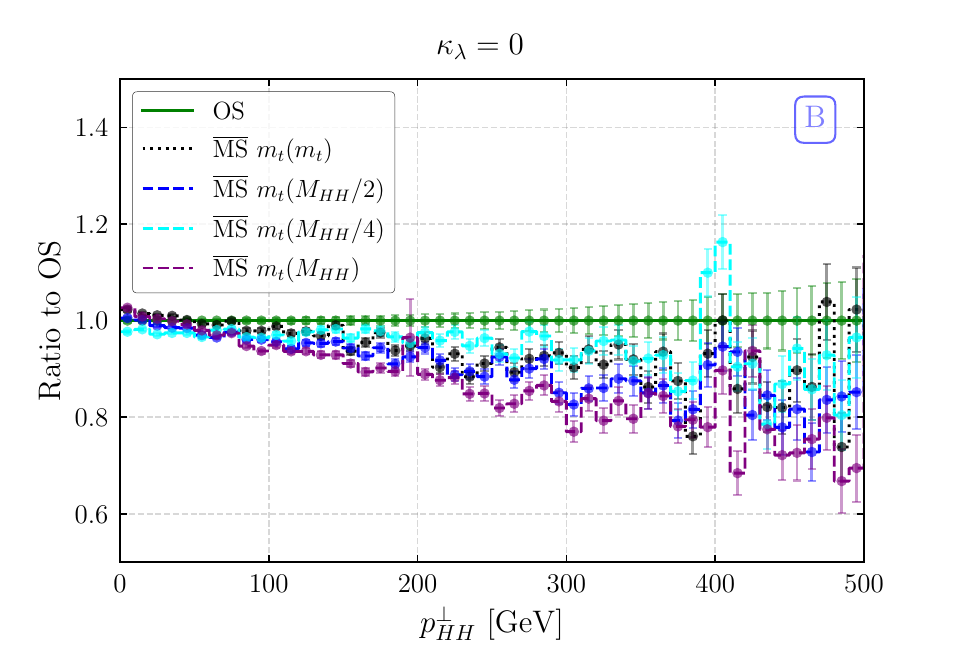}\\
  \includegraphics[width=0.5\textwidth]{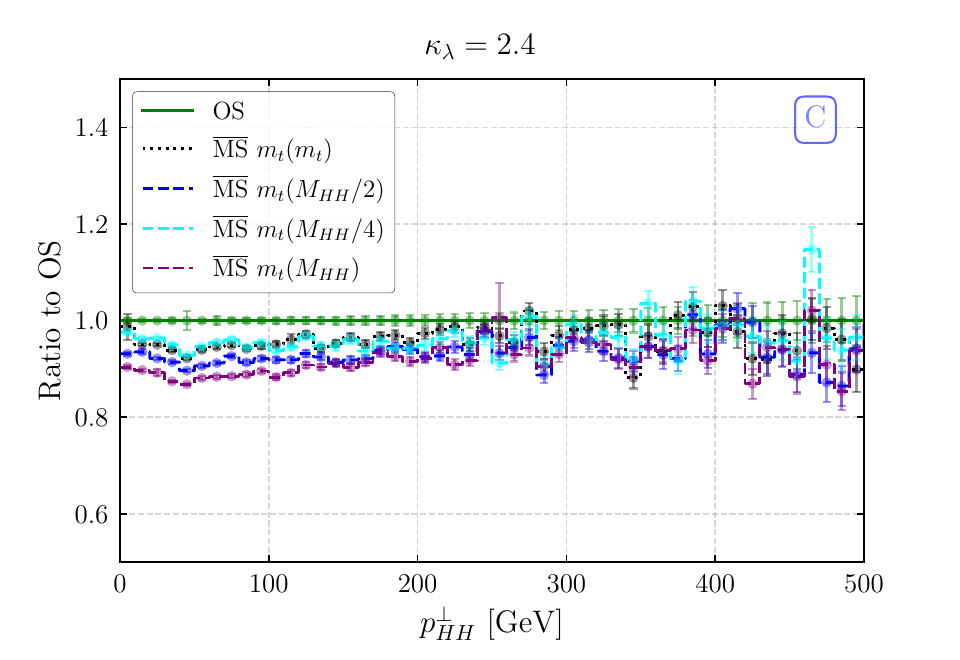}~\includegraphics[width=0.5\textwidth]{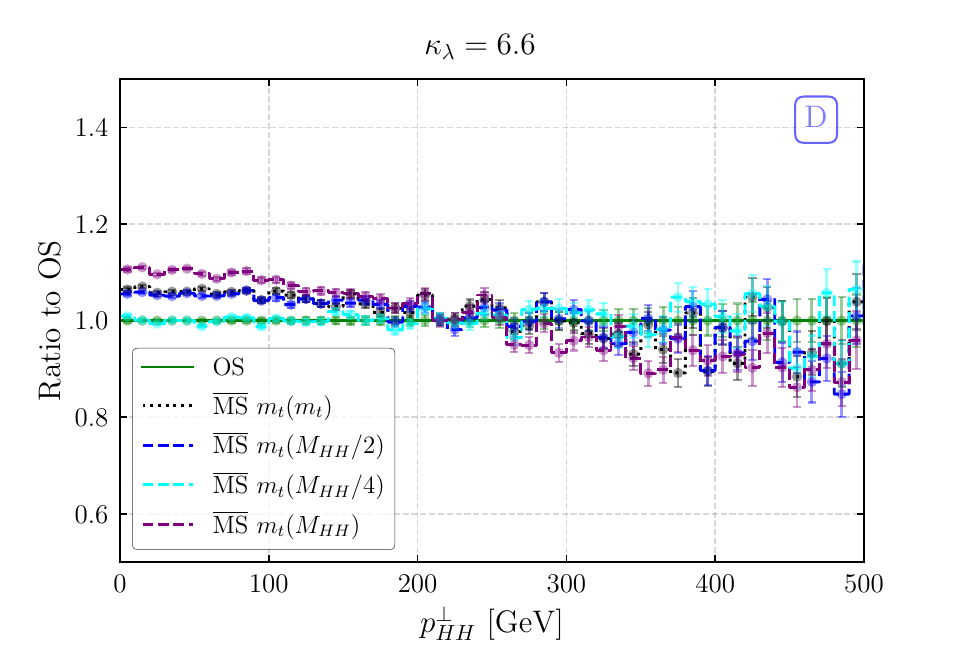}
  \caption{The transverse momentum distribution of the two-Higgs system
    for several values of $\ktre$ and different choices of the top-mass
    renormalization scheme: ratio between the
    $\MSbar$ predictions  and the  OS one for $\ktre = -0.6$ (\textbf{A}),
    $\ktre = 0$ (\textbf{B}), $\ktre = 2.4$ (\textbf{C}),
    $\ktre = 6.6$ (\textbf{D}).}
  \label{fig:ptHH-klambda}
\end{figure}

In fig.~\ref{fig:ptHH-klambda} we consider the transverse momentum
distribution of the two Higgs system. This figure has to be
compared with  fig.~\ref{fig:ptHH}~\textbf{B}. The cases $\ktre =-0.6$
(\textbf{A}) and $\ktre =0$ (\textbf{B}) are similar to the SM case although
with a slightly smaller spread among the $\MSbar$ predictions that also tend
to be closer to the OS one. Instead, the cases $\ktre = 2.4$ (\textbf{C})
and $\ktre= 6.6$ (\textbf{D})
show a scheme dependence quite similar for any value of $p^\perp_{HH}$
and relatively small especially in the $\ktre= 6.6$ case.

\begin{figure}
  \includegraphics[width=0.5\textwidth]{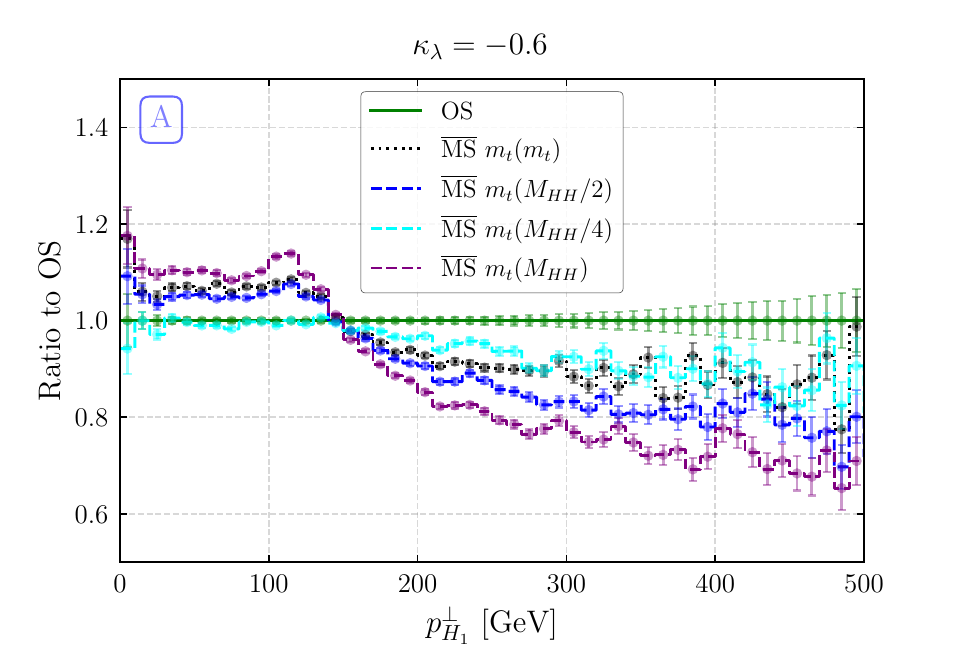}\includegraphics[width=0.5\textwidth]{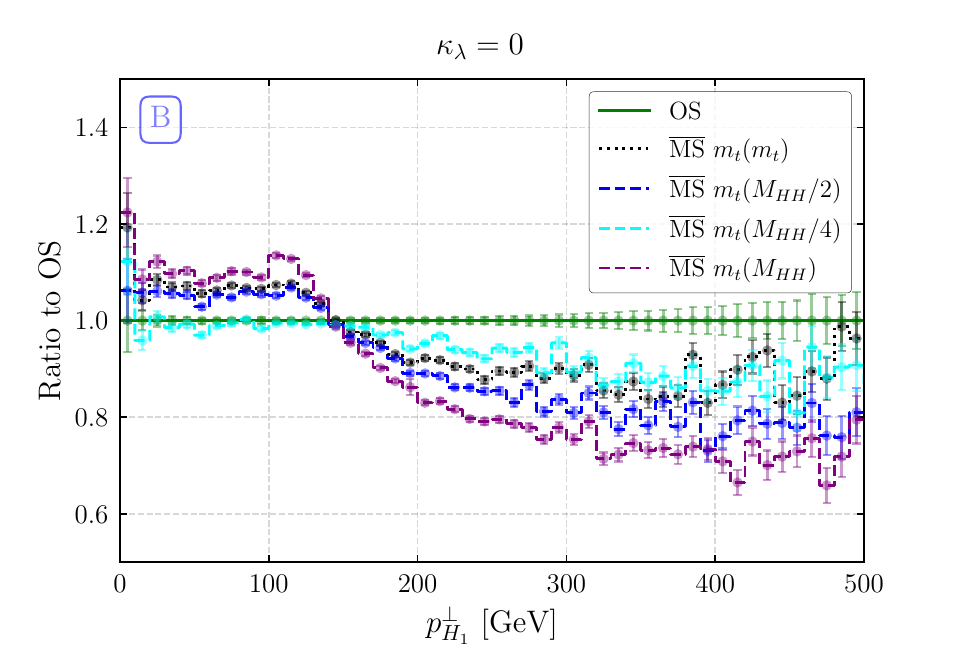}\\
  \includegraphics[width=0.5\textwidth]{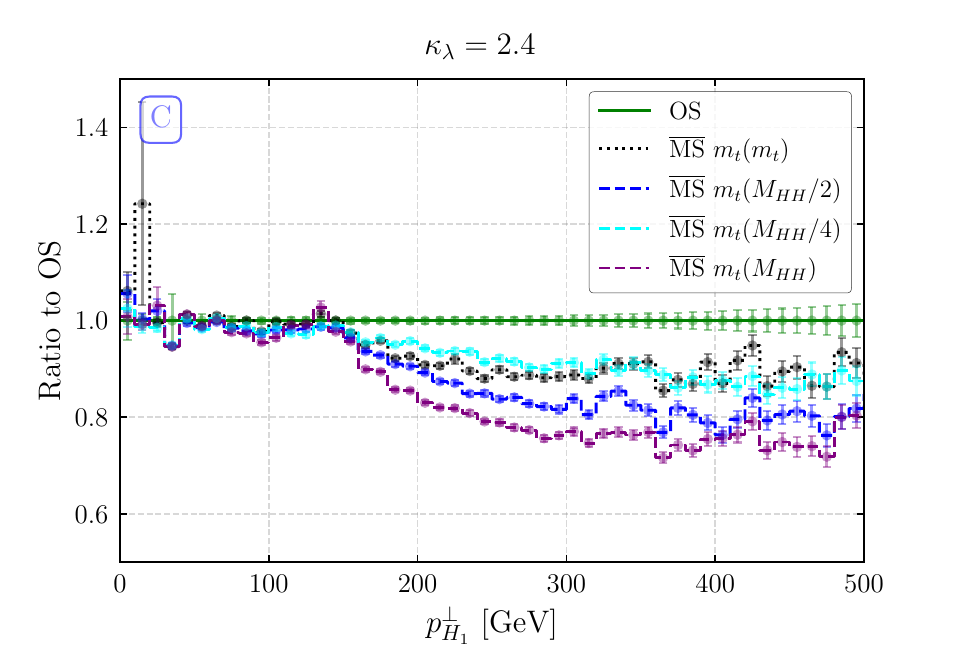}\includegraphics[width=0.5\textwidth]{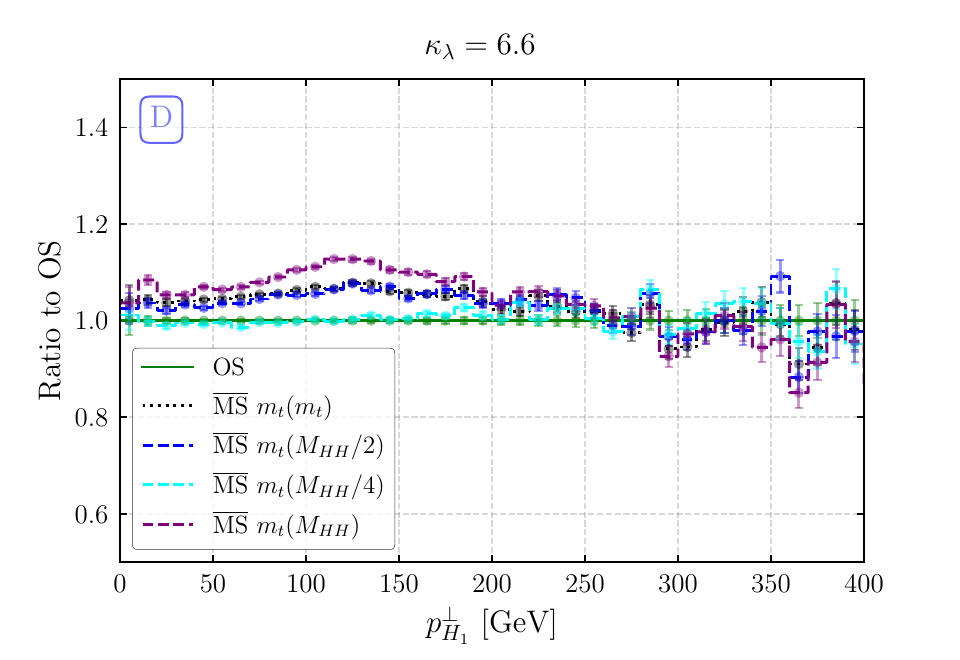}
  \caption{The transverse momentum distribution of the leading Higgs
    for several values of $\ktre$ and different choices of the top-mass
    renormalization scheme: ratio between the
    $\MSbar$ predictions  and the  OS one for $\ktre = -0.6$ (\textbf{A}),
    $\ktre = 0$ (\textbf{B}), $\ktre = 2.4$ (\textbf{C}),
    $\ktre = 6.6$ (\textbf{D}).}
  \label{fig:ptH1-klambda}
\end{figure}

Finally, fig.~\ref{fig:ptH1-klambda} presents the case of the transverse
momentum distribution of the leading Higgs to be compared with
fig.~\ref{fig:ptH1}~\textbf{B}. As for the previous observables
the cases $\ktre = -0.6$ (\textbf{A}) and $\ktre =0$ (\textbf{B}) are very
similar to the SM case. The case $\ktre = 2.4$ (\textbf{C}) shows a sensible
reduction of the
scheme dependence in the region $p^\perp_{HH} \lesssim 150$ GeV, while, as before,
for $\ktre =6.6$  (\textbf{D}) the scheme dependence is very reduced.

\section{Conclusions}
\label{sec:conc}
We have presented a new MC code for Higgs boson pair production at NLO in the
\POWHEGBOX approach. The main characteristic of this new code is the flexibility
both in the inputs parameters, including the Higgs trilinear coupling, and
in the choice of top-mass renormalization scheme. This is obtained employing
analytic results for the two-loop virtual contributions instead of numerical
grids.

Results are presented for the inclusive cross section and differential
observables, including parton shower, for several values of $\tril$ and
different choices of the top mass. We find, as a general trend, that going
from a LO result to a NLO one the top mass scheme dependence is reduced.
However, for large invariant mass of the Higgs-pair system or large transverse
momentum, the scheme dependence in the SM case, $\ktre =1 $, can be
significant, up to $20\%$. For other values of $\ktre$ we find similar results
but also cases where there is a more pronounced reduction of the top mass
scheme dependence.

We have compared our results with those of similar investigations present
in the literature \cite{Heinrich:2019bkc,Baglio:2020wgt}. We find a good
agreement for the SM case with all the results present in the literature.
For values of the Higgs trilinear coupling different from the SM one our results
for the inclusive cross section are in agreement with those of
the MC \ggHH, after the authors of that code corrected their evaluation of the
two-loop virtual contributions. With respect to the results of
Ref.\cite{Baglio:2020wgt} we found good  agreement only for $\ktre \le 1$.
Given the fact that in  our MC the cases $\ktre \neq 1$ are obtained just
assigning a value different from one to the parameter that multiplies the
``signal'' contribution we are quite confident in our result.

Finally we notice that our MC code is structured in such a way that can be
easily extended to include other beyond the SM effects besides the rescaling
of the Higgs trilinear coupling.

\section*{Acknowledgements}
G.D. and R.G. would like to thank their collaborators, Luigi
Bellafronte, Roberto Bonciani, Pier Paolo Giardino and Marco Vitti for
their contributions to our study of the NLO virtual corrections in
double Higgs production. We thank the authors of
Refs.~\cite{Heinrich:2019bkc, Baglio:2020wgt} for useful
communications. G.D.~thanks the Theory Division of CERN for support during the
initial part of this project. 
The work by R.G.~was supported by the PNRR CN1-Spoke 2.

\bibliographystyle{utphys}
\bibliography{gghh_nlo}

\providecommand{\href}[2]{#2}\begingroup\raggedright\begin{thebibliography}{10}

\bibitem{ATLAS:2012yve}
{\bfseries ATLAS} Collaboration, G.~Aad {\em et~al.}, ``{Observation of a new
  particle in the search for the Standard Model Higgs boson with the ATLAS
  detector at the LHC},''
  \href{http://dx.doi.org/10.1016/j.physletb.2012.08.020}{{\em Phys. Lett. B}
  {\bfseries 716} (2012) 1--29},
  \href{http://arxiv.org/abs/1207.7214}{{\ttfamily arXiv:1207.7214 [hep-ex]}}.

\bibitem{CMS:2012qbp}
{\bfseries CMS} Collaboration, S.~Chatrchyan {\em et~al.}, ``{Observation of a
  New Boson at a Mass of 125 GeV with the CMS Experiment at the LHC},''
  \href{http://dx.doi.org/10.1016/j.physletb.2012.08.021}{{\em Phys. Lett. B}
  {\bfseries 716} (2012) 30--61},
  \href{http://arxiv.org/abs/1207.7235}{{\ttfamily arXiv:1207.7235 [hep-ex]}}.

\bibitem{ATLAS:2022net}
{\bfseries ATLAS} Collaboration, ``{Measurement of the Higgs boson mass in the
  $H \rightarrow ZZ^* \rightarrow 4\ell$ decay channel using 139 fb$^{-1}$ of
  $\sqrt{s}=13$ TeV $pp$ collisions recorded by the ATLAS detector at the
  LHC},'' \href{http://arxiv.org/abs/2207.00320}{{\ttfamily arXiv:2207.00320
  [hep-ex]}}.

\bibitem{CMS:2020xrn}
{\bfseries CMS} Collaboration, A.~M. Sirunyan {\em et~al.}, ``{A measurement of
  the Higgs boson mass in the diphoton decay channel},''
  \href{http://dx.doi.org/10.1016/j.physletb.2020.135425}{{\em Phys. Lett. B}
  {\bfseries 805} (2020) 135425},
  \href{http://arxiv.org/abs/2002.06398}{{\ttfamily arXiv:2002.06398
  [hep-ex]}}.

\bibitem{DiMicco:2019ngk}
J.~Alison {\em et~al.}, ``{Higgs boson potential at colliders: Status and
  perspectives},'' \href{http://dx.doi.org/10.1016/j.revip.2020.100045}{{\em
  Rev. Phys.} {\bfseries 5} (2020) 100045},
  \href{http://arxiv.org/abs/1910.00012}{{\ttfamily arXiv:1910.00012
  [hep-ph]}}.

\bibitem{ATLAS:2022kbf}
{\bfseries ATLAS} Collaboration, ``{Constraining the Higgs boson self-coupling
  from single- and double-Higgs production with the ATLAS detector using $pp$
  collisions at $\sqrt{s}=13$ TeV},''
  \href{http://arxiv.org/abs/2211.01216}{{\ttfamily arXiv:2211.01216
  [hep-ex]}}.

\bibitem{CMS:2022dwd}
{\bfseries CMS} Collaboration, ``{A portrait of the Higgs boson by the CMS
  experiment ten years after the discovery},''
  \href{http://dx.doi.org/10.1038/s41586-022-04892-x}{{\em Nature} {\bfseries
  607} no.~7917, (2022) 60--68},
  \href{http://arxiv.org/abs/2207.00043}{{\ttfamily arXiv:2207.00043
  [hep-ex]}}.

\bibitem{Degrassi:2016wml}
G.~Degrassi, P.~P. Giardino, F.~Maltoni, and D.~Pagani, ``{Probing the Higgs
  self coupling via single Higgs production at the LHC},''
  \href{http://dx.doi.org/10.1007/JHEP12(2016)080}{{\em JHEP} {\bfseries 12}
  (2016) 080},
\href{http://arxiv.org/abs/1607.04251}{{\ttfamily arXiv:1607.04251 [hep-ph]}}.

\bibitem{Gorbahn:2016uoy}
M.~Gorbahn and U.~Haisch, ``{Indirect probes of the trilinear Higgs coupling:
  $gg \to h$ and $h \to \gamma \gamma$},''
  \href{http://dx.doi.org/10.1007/JHEP10(2016)094}{{\em JHEP} {\bfseries 10}
  (2016) 094},
\href{http://arxiv.org/abs/1607.03773}{{\ttfamily arXiv:1607.03773 [hep-ph]}}.

\bibitem{Bizon:2016wgr}
W.~Bizon, M.~Gorbahn, U.~Haisch, and G.~Zanderighi, ``{Constraints on the
  trilinear Higgs coupling from vector boson fusion and associated Higgs
  production at the LHC},''
  \href{http://dx.doi.org/10.1007/JHEP07(2017)083}{{\em JHEP} {\bfseries 07}
  (2017) 083},
\href{http://arxiv.org/abs/1610.05771}{{\ttfamily arXiv:1610.05771 [hep-ph]}}.

\bibitem{Maltoni:2017ims}
F.~Maltoni, D.~Pagani, A.~Shivaji, and X.~Zhao, ``{Trilinear Higgs coupling
  determination via single-Higgs differential measurements at the LHC},''
  \href{http://dx.doi.org/10.1140/epjc/s10052-017-5410-8}{{\em Eur. Phys. J.}
  {\bfseries C77} no.~12, (2017) 887},
\href{http://arxiv.org/abs/1709.08649}{{\ttfamily arXiv:1709.08649 [hep-ph]}}.

\bibitem{Maltoni:2018ttu}
F.~Maltoni, D.~Pagani, and X.~Zhao, ``{Constraining the Higgs self-couplings at
  e$^{+}$ e$^{-}$ colliders},''
  \href{http://dx.doi.org/10.1007/JHEP07(2018)087}{{\em JHEP} {\bfseries 07}
  (2018) 087},
\href{http://arxiv.org/abs/1802.07616}{{\ttfamily arXiv:1802.07616 [hep-ph]}}.

\bibitem{Gorbahn:2019lwq}
M.~Gorbahn and U.~Haisch, ``{Two-loop amplitudes for Higgs plus jet production
  involving a modified trilinear Higgs coupling},''
  \href{http://dx.doi.org/10.1007/JHEP04(2019)062}{{\em JHEP} {\bfseries 04}
  (2019) 062},
\href{http://arxiv.org/abs/1902.05480}{{\ttfamily arXiv:1902.05480 [hep-ph]}}.

\bibitem{Degrassi:2019yix}
G.~Degrassi and M.~Vitti, ``{The effect of an anomalous Higgs trilinear
  self-coupling on the $h \rightarrow \gamma \, Z$ decay},''
  \href{http://dx.doi.org/10.1140/epjc/s10052-020-7860-7}{{\em Eur. Phys. J. C}
  {\bfseries 80} no.~4, (2020) 307},
  \href{http://arxiv.org/abs/1912.06429}{{\ttfamily arXiv:1912.06429
  [hep-ph]}}.

\bibitem{Degrassi:2017ucl}
G.~Degrassi, M.~Fedele, and P.~P. Giardino, ``{Constraints on the trilinear
  Higgs self coupling from precision observables},''
  \href{http://dx.doi.org/10.1007/JHEP04(2017)155}{{\em JHEP} {\bfseries 04}
  (2017) 155},
\href{http://arxiv.org/abs/1702.01737}{{\ttfamily arXiv:1702.01737 [hep-ph]}}.

\bibitem{Kribs:2017znd}
G.~D. Kribs, A.~Maier, H.~Rzehak, M.~Spannowsky, and P.~Waite, ``{Electroweak
  oblique parameters as a probe of the trilinear Higgs boson
  self-interaction},'' \href{http://dx.doi.org/10.1103/PhysRevD.95.093004}{{\em
  Phys. Rev.} {\bfseries D95} no.~9, (2017) 093004},
\href{http://arxiv.org/abs/1702.07678}{{\ttfamily arXiv:1702.07678 [hep-ph]}}.

\bibitem{Degrassi:2021uik}
G.~Degrassi, B.~Di~Micco, P.~P. Giardino, and E.~Rossi, ``{Higgs boson
  self-coupling constraints from single Higgs, double Higgs and Electroweak
  measurements},'' \href{http://dx.doi.org/10.1016/j.physletb.2021.136307}{{\em
  Phys. Lett. B} {\bfseries 817} (2021) 136307},
  \href{http://arxiv.org/abs/2102.07651}{{\ttfamily arXiv:2102.07651
  [hep-ph]}}.

\bibitem{DiVita:2017eyz}
S.~Di~Vita, C.~Grojean, G.~Panico, M.~Riembau, and T.~Vantalon, ``{A global
  view on the Higgs self-coupling},''
  \href{http://dx.doi.org/10.1007/JHEP09(2017)069}{{\em JHEP} {\bfseries 09}
  (2017) 069}, \href{http://arxiv.org/abs/1704.01953}{{\ttfamily
  arXiv:1704.01953 [hep-ph]}}.

\bibitem{Alasfar:2022zyr}
L.~Alasfar, J.~de~Blas, and R.~Gr\"ober, ``{Higgs probes of top quark contact
  interactions and their interplay with the Higgs self-coupling},''
  \href{http://dx.doi.org/10.1007/JHEP05(2022)111}{{\em JHEP} {\bfseries 05}
  (2022) 111}, \href{http://arxiv.org/abs/2202.02333}{{\ttfamily
  arXiv:2202.02333 [hep-ph]}}.

\bibitem{Glover:1987nx}
E.~W.~N. Glover and J.~J. van~der Bij, ``{Higgs Boson Pair Production via Gluon
  Fusion},''
\href{http://dx.doi.org/10.1016/0550-3213(88)90083-1}{{\em Nucl. Phys.}
  {\bfseries B309} (1988) 282--294}.

\bibitem{Dicus:1987ic}
D.~A. Dicus, C.~Kao, and S.~S.~D. Willenbrock, ``{Higgs Boson Pair Production
  From Gluon Fusion},''
  \href{http://dx.doi.org/10.1016/0370-2693(88)90202-X}{{\em Phys. Lett. B}
  {\bfseries 203} 457--461}.

\bibitem{Plehn:1996wb}
T.~Plehn, M.~Spira, and P.~M. Zerwas, ``{Pair production of neutral Higgs
  particles in gluon-gluon coll isions},''
  \href{http://dx.doi.org/10.1016/0550-3213(96)00418-X}{{\em Nucl. Phys. B}
  {\bfseries 479} (1996) 46--64},
  \href{http://arxiv.org/abs/hep-ph/9603205}{{\ttfamily arXiv:hep-ph/9603205}}.
  [Erratum: Nucl.Phys.B 531, 655--655 (1998)].

\bibitem{Dawson:1998py}
S.~Dawson, S.~Dittmaier, and M.~Spira, ``{Neutral Higgs boson pair production
  at hadron colliders: QCD corrections},''
  \href{http://dx.doi.org/10.1103/PhysRevD.58.115012}{{\em Phys. Rev.}
  {\bfseries D58} (1998) 115012},
\href{http://arxiv.org/abs/hep-ph/9805244}{{\ttfamily arXiv:hep-ph/9805244
  [hep-ph]}}.

\bibitem{Grigo:2013rya}
J.~Grigo, J.~Hoff, K.~Melnikov, and M.~Steinhauser, ``{On the Higgs boson pair
  production at the LHC},''
  \href{http://dx.doi.org/10.1016/j.nuclphysb.2013.06.024}{{\em Nucl. Phys.}
  {\bfseries B875} (2013) 1--17},
\href{http://arxiv.org/abs/1305.7340}{{\ttfamily arXiv:1305.7340 [hep-ph]}}.

\bibitem{Grigo:2015dia}
J.~Grigo, J.~Hoff, and M.~Steinhauser, ``{Higgs boson pair production: top
  quark mass effects at NLO and NNLO},''
  \href{http://dx.doi.org/10.1016/j.nuclphysb.2015.09.012}{{\em Nucl. Phys.}
  {\bfseries B900} (2015) 412--430},
\href{http://arxiv.org/abs/1508.00909}{{\ttfamily arXiv:1508.00909 [hep-ph]}}.

\bibitem{Degrassi:2016vss}
G.~Degrassi, P.~P. Giardino, and R.~Gr{\"o}ber, ``{On the two-loop virtual QCD
  corrections to Higgs boson pair production in the Standard Model},''
  \href{http://dx.doi.org/10.1140/epjc/s10052-016-4256-9}{{\em Eur. Phys. J.}
  {\bfseries C76} no.~7, (2016) 411},
\href{http://arxiv.org/abs/1603.00385}{{\ttfamily arXiv:1603.00385 [hep-ph]}}.

\bibitem{Frederix:2014hta}
R.~Frederix, S.~Frixione, V.~Hirschi, F.~Maltoni, O.~Mattelaer, P.~Torrielli,
  E.~Vryonidou, and M.~Zaro, ``{Higgs pair production at the LHC with NLO and
  parton-shower effects},''
  \href{http://dx.doi.org/10.1016/j.physletb.2014.03.026}{{\em Phys. Lett. B}
  {\bfseries 732} (2014) 142--149},
  \href{http://arxiv.org/abs/1401.7340}{{\ttfamily arXiv:1401.7340 [hep-ph]}}.

\bibitem{Maltoni:2014eza}
F.~Maltoni, E.~Vryonidou, and M.~Zaro, ``{Top-quark mass effects in double and
  triple Higgs production in gluon-gluon fusion at NLO},''
  \href{http://dx.doi.org/10.1007/JHEP11(2014)079}{{\em JHEP} {\bfseries 11}
  (2014) 079}, \href{http://arxiv.org/abs/1408.6542}{{\ttfamily arXiv:1408.6542
  [hep-ph]}}.

\bibitem{Borowka:2016ehy}
S.~Borowka, N.~Greiner, G.~Heinrich, S.~Jones, M.~Kerner, J.~Schlenk,
  U.~Schubert, and T.~Zirke, ``{Higgs Boson Pair Production in Gluon Fusion at
  Next-to-Leading Order with Full Top-Quark Mass Dependence},''
  \href{http://dx.doi.org/10.1103/PhysRevLett.117.079901,
  10.1103/PhysRevLett.117.012001}{{\em Phys. Rev. Lett.} {\bfseries 117} no.~1,
  (2016) 012001}, \href{http://arxiv.org/abs/1604.06447}{{\ttfamily
  arXiv:1604.06447 [hep-ph]}}.
[Erratum: Phys. Rev. Lett.117,no.7,079901(2016)].

\bibitem{Borowka:2016ypz}
S.~Borowka, N.~Greiner, G.~Heinrich, S.~P. Jones, M.~Kerner, J.~Schlenk, and
  T.~Zirke, ``{Full top quark mass dependence in Higgs boson pair production at
  NLO},'' \href{http://dx.doi.org/10.1007/JHEP10(2016)107}{{\em JHEP}
  {\bfseries 10} (2016) 107},
\href{http://arxiv.org/abs/1608.04798}{{\ttfamily arXiv:1608.04798 [hep-ph]}}.

\bibitem{Baglio:2018lrj}
J.~Baglio, F.~Campanario, S.~Glaus, M.~M\"uhlleitner, M.~Spira, and
  J.~Streicher, ``{Gluon fusion into Higgs pairs at NLO QCD and the top mass
  scheme},'' \href{http://dx.doi.org/10.1140/epjc/s10052-019-6973-3}{{\em Eur.
  Phys. J. C} {\bfseries 79} no.~6, (2019) 459},
  \href{http://arxiv.org/abs/1811.05692}{{\ttfamily arXiv:1811.05692
  [hep-ph]}}.

\bibitem{Baglio:2020ini}
J.~Baglio, F.~Campanario, S.~Glaus, M.~M\"uhlleitner, J.~Ronca, M.~Spira, and
  J.~Streicher, ``{Higgs-Pair Production via Gluon Fusion at Hadron Colliders:
  NLO QCD Corrections},'' \href{http://dx.doi.org/10.1007/JHEP04(2020)181}{{\em
  JHEP} {\bfseries 04} (2020) 181},
  \href{http://arxiv.org/abs/2003.03227}{{\ttfamily arXiv:2003.03227
  [hep-ph]}}.

\bibitem{Bonciani:2018omm}
R.~Bonciani, G.~Degrassi, P.~P. Giardino, and R.~Gr\"ober, ``{Analytical Method
  for Next-to-Leading-Order QCD Corrections to Double-Higgs Production},''
  \href{http://dx.doi.org/10.1103/PhysRevLett.121.162003}{{\em Phys. Rev.
  Lett.} {\bfseries 121} no.~16, (2018) 162003},
  \href{http://arxiv.org/abs/1806.11564}{{\ttfamily arXiv:1806.11564
  [hep-ph]}}.

\bibitem{Davies:2018ood}
J.~Davies, G.~Mishima, M.~Steinhauser, and D.~Wellmann, ``{Double-Higgs boson
  production in the high-energy limit: planar master integrals},''
  \href{http://dx.doi.org/10.1007/JHEP03(2018)048}{{\em JHEP} {\bfseries 03}
  (2018) 048},
\href{http://arxiv.org/abs/1801.09696}{{\ttfamily arXiv:1801.09696 [hep-ph]}}.

\bibitem{Davies:2018qvx}
J.~Davies, G.~Mishima, M.~Steinhauser, and D.~Wellmann, ``{Double Higgs boson
  production at NLO in the high-energy limit: complete analytic results},''
  \href{http://dx.doi.org/10.1007/JHEP01(2019)176}{{\em JHEP} {\bfseries 01}
  (2019) 176}, \href{http://arxiv.org/abs/1811.05489}{{\ttfamily
  arXiv:1811.05489 [hep-ph]}}.

\bibitem{Wang:2020nnr}
G.~Wang, Y.~Wang, X.~Xu, Y.~Xu, and L.~L. Yang, ``{Efficient computation of
  two-loop amplitudes for Higgs boson pair production},''
  \href{http://dx.doi.org/10.1103/PhysRevD.104.L051901}{{\em Phys. Rev. D}
  {\bfseries 104} no.~5, (2021) L051901},
  \href{http://arxiv.org/abs/2010.15649}{{\ttfamily arXiv:2010.15649
  [hep-ph]}}.

\bibitem{Davies:2019dfy}
J.~Davies, G.~Heinrich, S.~P. Jones, M.~Kerner, G.~Mishima, M.~Steinhauser, and
  D.~Wellmann, ``{Double Higgs boson production at NLO: combining the exact
  numerical result and high-energy expansion},''
  \href{http://dx.doi.org/10.1007/JHEP11(2019)024}{{\em JHEP} {\bfseries 11}
  (2019) 024}, \href{http://arxiv.org/abs/1907.06408}{{\ttfamily
  arXiv:1907.06408 [hep-ph]}}.

\bibitem{Bellafronte:2022jmo}
L.~Bellafronte, G.~Degrassi, P.~P. Giardino, R.~Gr\"ober, and M.~Vitti,
  ``{Gluon fusion production at NLO: merging the transverse momentum and the
  high-energy expansions},''
  \href{http://dx.doi.org/10.1007/JHEP07(2022)069}{{\em JHEP} {\bfseries 07}
  (2022) 069}, \href{http://arxiv.org/abs/2202.12157}{{\ttfamily
  arXiv:2202.12157 [hep-ph]}}.

\bibitem{Heinrich:2017kxx}
G.~Heinrich, S.~P. Jones, M.~Kerner, G.~Luisoni, and E.~Vryonidou, ``{NLO
  predictions for Higgs boson pair production with full top quark mass
  dependence matched to parton showers},''
  \href{http://dx.doi.org/10.1007/JHEP08(2017)088}{{\em JHEP} {\bfseries 08}
  (2017) 088},
\href{http://arxiv.org/abs/1703.09252}{{\ttfamily arXiv:1703.09252 [hep-ph]}}.

\bibitem{Jones:2017giv}
S.~Jones and S.~Kuttimalai, ``{Parton Shower and NLO-Matching uncertainties in
  Higgs Boson Pair Production},''
  \href{http://dx.doi.org/10.1007/JHEP02(2018)176}{{\em JHEP} {\bfseries 02}
  (2018) 176}, \href{http://arxiv.org/abs/1711.03319}{{\ttfamily
  arXiv:1711.03319 [hep-ph]}}.

\bibitem{deFlorian:2013uza}
D.~de~Florian and J.~Mazzitelli, ``{Two-loop virtual corrections to Higgs pair
  production},'' \href{http://dx.doi.org/10.1016/j.physletb.2013.06.046}{{\em
  Phys. Lett. B} {\bfseries 724} (2013) 306--309},
  \href{http://arxiv.org/abs/1305.5206}{{\ttfamily arXiv:1305.5206 [hep-ph]}}.

\bibitem{deFlorian:2013jea}
D.~de~Florian and J.~Mazzitelli, ``{Higgs Boson Pair Production at
  Next-to-Next-to-Leading Order in QCD},''
  \href{http://dx.doi.org/10.1103/PhysRevLett.111.201801}{{\em Phys. Rev.
  Lett.} {\bfseries 111} (2013) 201801},
  \href{http://arxiv.org/abs/1309.6594}{{\ttfamily arXiv:1309.6594 [hep-ph]}}.

\bibitem{Grigo:2014jma}
J.~Grigo, K.~Melnikov, and M.~Steinhauser, ``{Virtual corrections to Higgs
  boson pair production in the large top quark mass limit},''
  \href{http://dx.doi.org/10.1016/j.nuclphysb.2014.09.003}{{\em Nucl. Phys. B}
  {\bfseries 888} (2014) 17--29},
  \href{http://arxiv.org/abs/1408.2422}{{\ttfamily arXiv:1408.2422 [hep-ph]}}.

\bibitem{deFlorian:2016uhr}
D.~de~Florian, M.~Grazzini, C.~Hanga, S.~Kallweit, J.~M. Lindert,
  P.~Maierh\"ofer, J.~Mazzitelli, and D.~Rathlev, ``{Differential Higgs Boson
  Pair Production at Next-to-Next-to-Leading Order in QCD},''
  \href{http://dx.doi.org/10.1007/JHEP09(2016)151}{{\em JHEP} {\bfseries 09}
  (2016) 151}, \href{http://arxiv.org/abs/1606.09519}{{\ttfamily
  arXiv:1606.09519 [hep-ph]}}.

\bibitem{Grazzini:2018bsd}
M.~Grazzini, G.~Heinrich, S.~Jones, S.~Kallweit, M.~Kerner, J.~M. Lindert, and
  J.~Mazzitelli, ``{Higgs boson pair production at NNLO with top quark mass
  effects},'' \href{http://dx.doi.org/10.1007/JHEP05(2018)059}{{\em JHEP}
  {\bfseries 05} (2018) 059}, \href{http://arxiv.org/abs/1803.02463}{{\ttfamily
  arXiv:1803.02463 [hep-ph]}}.

\bibitem{Chen:2019lzz}
L.-B. Chen, H.~T. Li, H.-S. Shao, and J.~Wang, ``{Higgs boson pair production
  via gluon fusion at N$^3$LO in QCD},''
  \href{http://dx.doi.org/10.1016/j.physletb.2020.135292}{{\em Phys. Lett. B}
  {\bfseries 803} (2020) 135292},
  \href{http://arxiv.org/abs/1909.06808}{{\ttfamily arXiv:1909.06808
  [hep-ph]}}.

\bibitem{Chen:2019fhs}
L.-B. Chen, H.~T. Li, H.-S. Shao, and J.~Wang, ``{The gluon-fusion production
  of Higgs boson pair: N$^3$LO QCD corrections and top-quark mass effects},''
  \href{http://dx.doi.org/10.1007/JHEP03(2020)072}{{\em JHEP} {\bfseries 03}
  (2020) 072}, \href{http://arxiv.org/abs/1912.13001}{{\ttfamily
  arXiv:1912.13001 [hep-ph]}}.

\bibitem{Frixione:2007vw}
S.~Frixione, P.~Nason, and C.~Oleari, ``{Matching NLO QCD computations with
  Parton Shower simulations: the POWHEG method},''
  \href{http://dx.doi.org/10.1088/1126-6708/2007/11/070}{{\em JHEP} {\bfseries
  11} (2007) 070}, \href{http://arxiv.org/abs/0709.2092}{{\ttfamily
  arXiv:0709.2092 [hep-ph]}}.

\bibitem{Alioli:2010xd}
S.~Alioli, P.~Nason, C.~Oleari, and E.~Re, ``{A general framework for
  implementing NLO calculations in shower Monte Carlo programs: the POWHEG
  BOX},'' \href{http://dx.doi.org/10.1007/JHEP06(2010)043}{{\em JHEP}
  {\bfseries 06} (2010) 043}, \href{http://arxiv.org/abs/1002.2581}{{\ttfamily
  arXiv:1002.2581 [hep-ph]}}.

\bibitem{Heinrich:2019bkc}
G.~Heinrich, S.~P. Jones, M.~Kerner, G.~Luisoni, and L.~Scyboz, ``{Probing the
  trilinear Higgs boson coupling in di-Higgs production at NLO QCD including
  parton shower effects},''
  \href{http://dx.doi.org/10.1007/JHEP06(2019)066}{{\em JHEP} {\bfseries 06}
  (2019) 066}, \href{http://arxiv.org/abs/1903.08137}{{\ttfamily
  arXiv:1903.08137 [hep-ph]}}.

\bibitem{Baglio:2020wgt}
J.~Baglio, F.~Campanario, S.~Glaus, M.~M\"uhlleitner, J.~Ronca, and M.~Spira,
  ``{$gg\to HH$ : Combined uncertainties},''
  \href{http://dx.doi.org/10.1103/PhysRevD.103.056002}{{\em Phys. Rev. D}
  {\bfseries 103} no.~5, (2021) 056002},
  \href{http://arxiv.org/abs/2008.11626}{{\ttfamily arXiv:2008.11626
  [hep-ph]}}.

\bibitem{Passarino:1978jh}
G.~Passarino and M.~J.~G. Veltman, ``{One Loop Corrections for $e^+ \,e^-$
  Annihilation Into $\mu^+\, \mu^-$ in the Weinberg Model},''
  \href{http://dx.doi.org/10.1016/0550-3213(79)90234-7}{{\em Nucl. Phys. B}
  {\bfseries 160} (1979) 151--207}.

\bibitem{Denner:2014gla}
A.~Denner, S.~Dittmaier, and L.~Hofer, ``{COLLIER - A fortran-library for
  one-loop integrals},'' \href{http://dx.doi.org/10.22323/1.211.0071}{{\em PoS}
  {\bfseries LL2014} (2014) 071},
  \href{http://arxiv.org/abs/1407.0087}{{\ttfamily arXiv:1407.0087 [hep-ph]}}.

\bibitem{Aglietti:2006tp}
U.~Aglietti, R.~Bonciani, G.~Degrassi, and A.~Vicini, ``{Analytic Results for
  Virtual QCD Corrections to Higgs Production and Decay},''
  \href{http://dx.doi.org/10.1088/1126-6708/2007/01/021}{{\em JHEP} {\bfseries
  01} (2007) 021},
\href{http://arxiv.org/abs/hep-ph/0611266}{{\ttfamily arXiv:hep-ph/0611266
  [hep-ph]}}.

\bibitem{Anastasiou:2006hc}
C.~Anastasiou, S.~Beerli, S.~Bucherer, A.~Daleo, and Z.~Kunszt, ``{Two-loop
  amplitudes and master integrals for the production of a Higgs boson via a
  massive quark and a scalar-quark loop},''
  \href{http://dx.doi.org/10.1088/1126-6708/2007/01/082}{{\em JHEP} {\bfseries
  01} (2007) 082},
\href{http://arxiv.org/abs/hep-ph/0611236}{{\ttfamily arXiv:hep-ph/0611236
  [hep-ph]}}.

\bibitem{Lee:2013mka}
R.~N. Lee, ``{LiteRed 1.4: a powerful tool for reduction of multiloop
  integrals},'' \href{http://dx.doi.org/10.1088/1742-6596/523/1/012059}{{\em J.
  Phys. Conf. Ser.} {\bfseries 523} (2014) 012059},
\href{http://arxiv.org/abs/1310.1145}{{\ttfamily arXiv:1310.1145 [hep-ph]}}.

\bibitem{Lee:2012cn}
R.~N. Lee, ``{Presenting LiteRed: a tool for the Loop InTEgrals REDuction},''
  \href{http://arxiv.org/abs/1212.2685}{{\ttfamily arXiv:1212.2685 [hep-ph]}}.

\bibitem{vonManteuffel:2017hms}
A.~von Manteuffel and L.~Tancredi, ``{A non-planar two-loop three-point
  function beyond multiple polylogarithms},''
  \href{http://dx.doi.org/10.1007/JHEP06(2017)127}{{\em JHEP} {\bfseries 06}
  (2017) 127},
\href{http://arxiv.org/abs/1701.05905}{{\ttfamily arXiv:1701.05905 [hep-ph]}}.

\bibitem{GitHub}
\url{https://github.com/mppmu/hhgrid}.

\bibitem{Naterop:2019xaf}
L.~Naterop, A.~Signer, and Y.~Ulrich, ``{handyG \textemdash{}Rapid numerical
  evaluation of generalised polylogarithms in Fortran},''
  \href{http://dx.doi.org/10.1016/j.cpc.2020.107165}{{\em Comput. Phys.
  Commun.} {\bfseries 253} (2020) 107165},
  \href{http://arxiv.org/abs/1909.01656}{{\ttfamily arXiv:1909.01656
  [hep-ph]}}.

\bibitem{Bonciani:2018uvv}
R.~Bonciani, G.~Degrassi, P.~P. Giardino, and R.~Gr\"ober, ``{A Numerical
  Routine for the Crossed Vertex Diagram with a Massive-Particle Loop},''
  \href{http://dx.doi.org/10.1016/j.cpc.2019.03.014}{{\em Comput. Phys.
  Commun.} {\bfseries 241} (2019) 122--131},
  \href{http://arxiv.org/abs/1812.02698}{{\ttfamily arXiv:1812.02698
  [hep-ph]}}.

\bibitem{Melnikov:2000qh}
K.~Melnikov and T.~v. Ritbergen, ``{The Three loop relation between the MS-bar
  and the pole quark masses},''
  \href{http://dx.doi.org/10.1016/S0370-2693(00)00507-4}{{\em Phys. Lett. B}
  {\bfseries 482} (2000) 99--108},
  \href{http://arxiv.org/abs/hep-ph/9912391}{{\ttfamily arXiv:hep-ph/9912391}}.

\bibitem{Carena:1999py}
M.~Carena, D.~Garcia, U.~Nierste, and C.~E.~M. Wagner, ``{Effective Lagrangian
  for the $\bar{t} b H^{+}$ interaction in the MSSM and charged Higgs
  phenomenology},'' \href{http://dx.doi.org/10.1016/S0550-3213(00)00146-2}{{\em
  Nucl. Phys. B} {\bfseries 577} (2000) 88--120},
  \href{http://arxiv.org/abs/hep-ph/9912516}{{\ttfamily arXiv:hep-ph/9912516}}.

\bibitem{Hirschi:2011pa}
V.~Hirschi, R.~Frederix, S.~Frixione, M.~V. Garzelli, F.~Maltoni, and
  R.~Pittau, ``{Automation of one-loop QCD corrections},''
  \href{http://dx.doi.org/10.1007/JHEP05(2011)044}{{\em JHEP} {\bfseries 05}
  (2011) 044}, \href{http://arxiv.org/abs/1103.0621}{{\ttfamily arXiv:1103.0621
  [hep-ph]}}.

\bibitem{Bagnaschi:2015qta}
E.~Bagnaschi and A.~Vicini, ``{The Higgs transverse momentum distribution in
  gluon fusion as a multiscale problem},''
  \href{http://dx.doi.org/10.1007/JHEP01(2016)056}{{\em JHEP} {\bfseries 01}
  (2016) 056}, \href{http://arxiv.org/abs/1505.00735}{{\ttfamily
  arXiv:1505.00735 [hep-ph]}}.

\bibitem{Bagnaschi:2015bop}
E.~Bagnaschi, R.~V. Harlander, H.~Mantler, A.~Vicini, and M.~Wiesemann,
  ``{Resummation ambiguities in the Higgs transverse-momentum spectrum in the
  Standard Model and beyond},''
  \href{http://dx.doi.org/10.1007/JHEP01(2016)090}{{\em JHEP} {\bfseries 01}
  (2016) 090}, \href{http://arxiv.org/abs/1510.08850}{{\ttfamily
  arXiv:1510.08850 [hep-ph]}}.

\bibitem{Alasfar:2023xpc}
L.~Alasfar {\em et~al.}, ``{Effective Field Theory descriptions of Higgs boson
  pair production},'' \href{http://arxiv.org/abs/2304.01968}{{\ttfamily
  arXiv:2304.01968 [hep-ph]}}.

\bibitem{GoSam:2014iqq}
{\bfseries GoSam} Collaboration, G.~Cullen {\em et~al.},
  ``{G$\scriptsize{O}$S$\scriptsize{AM}$-2.0: a tool for automated one-loop
  calculations within the Standard Model and beyond},''
  \href{http://dx.doi.org/10.1140/epjc/s10052-014-3001-5}{{\em Eur. Phys. J. C}
  {\bfseries 74} no.~8, (2014) 3001},
  \href{http://arxiv.org/abs/1404.7096}{{\ttfamily arXiv:1404.7096 [hep-ph]}}.

\bibitem{Cullen:2011ac}
{\bfseries GoSam} Collaboration, G.~Cullen, N.~Greiner, G.~Heinrich,
  G.~Luisoni, P.~Mastrolia, G.~Ossola, T.~Reiter, and F.~Tramontano,
  ``{Automated One-Loop Calculations with GoSam},''
  \href{http://dx.doi.org/10.1140/epjc/s10052-012-1889-1}{{\em Eur. Phys. J. C}
  {\bfseries 72} (2012) 1889}, \href{http://arxiv.org/abs/1111.2034}{{\ttfamily
  arXiv:1111.2034 [hep-ph]}}.

\bibitem{NNPDF:2017mvq}
{\bfseries NNPDF} Collaboration, R.~D. Ball {\em et~al.}, ``{Parton
  distributions from high-precision collider data},''
  \href{http://dx.doi.org/10.1140/epjc/s10052-017-5199-5}{{\em Eur. Phys. J. C}
  {\bfseries 77} no.~10, (2017) 663},
  \href{http://arxiv.org/abs/1706.00428}{{\ttfamily arXiv:1706.00428
  [hep-ph]}}.

\bibitem{Sjostrand:2007gs}
T.~Sjostrand, S.~Mrenna, and P.~Z. Skands, ``{A Brief Introduction to PYTHIA
  8.1},'' \href{http://dx.doi.org/10.1016/j.cpc.2008.01.036}{{\em Comput. Phys.
  Commun.} {\bfseries 178} (2008) 852--867},
  \href{http://arxiv.org/abs/0710.3820}{{\ttfamily arXiv:0710.3820 [hep-ph]}}.

\bibitem{Sjostrand:2014zea}
T.~Sj\"ostrand, S.~Ask, J.~R. Christiansen, R.~Corke, N.~Desai, P.~Ilten,
  S.~Mrenna, S.~Prestel, C.~O. Rasmussen, and P.~Z. Skands, ``{An introduction
  to PYTHIA 8.2},'' \href{http://dx.doi.org/10.1016/j.cpc.2015.01.024}{{\em
  Comput. Phys. Commun.} {\bfseries 191} (2015) 159--177},
  \href{http://arxiv.org/abs/1410.3012}{{\ttfamily arXiv:1410.3012 [hep-ph]}}.

\end{thebibliography}\endgroup

\end{document}